\documentclass[preprint,12pt]{elsarticle}
\usepackage{setspace}
\usepackage{mathrsfs}
\usepackage{amsmath}
\usepackage{booktabs}
\usepackage{amssymb}
\usepackage{accents}
\usepackage{statex}
\usepackage{enumerate}
\usepackage{graphicx}
\usepackage{subfigure}
\usepackage{indentfirst}
\usepackage{geometry}
\usepackage{array}
\usepackage{booktabs}
\usepackage{tabularx}
\usepackage{multicol}
\usepackage{multirow}
\newtheorem{thm}{Theorem}[section]
\newtheorem{lem}{Lemma}[section]

\newtheorem{cor}{Corollary}[section]
\newtheorem{pro}{Proof}
\newtheorem{defn}{Definition}[section]
\newcommand{\PreserveBackslash}[1]{\let\temp=\\#1\let\\=\temp}
\newcommand\relphantom[1]{\mathrel{\phantom{#1}}}
\newcolumntype{C}[1]{>{\PreserveBackslash\centering}p{#1}}
\newcolumntype{R}[1]{>{\PreserveBackslash\raggedleft}p{#1}}
\newcolumntype{L}[1]{>{\PreserveBackslash\raggedright}p{#1}}
\journal{Journal}
\begin{document}
\begin{frontmatter}
\title{$N$-dimensional Heisenberg's uncertainty principle for fractional Fourier transform}
\tnotetext[mytitlenote]{The work was supported by the Startup Foundation for Introducing Talent of NUIST (Grant 2019r024).}
\author[label1]{Zhichao Zhang\corref{cor1}}
\ead{zzc910731@163.com, zhichaozhang2013scu@gmail.com.}
\cortext[cor1]{Corresponding author; Tel: +86-13376073017.}
\address[label1]{School of Mathematics and Statistics, Nanjing University of Information Science \& Technology, Nanjing 210044, China}
\begin{abstract}
\noindent A sharper uncertainty inequality which exhibits a lower bound larger than that in the classical $N$-dimensional Heisenberg's uncertainty principle is obtained, and extended from $N$-dimensional Fourier transform domain to two $N$-dimensional fractional Fourier transform domains. The conditions that reach the equality relation of the uncertainty inequalities are deduced. Example and simulation are performed to illustrate that the newly derived uncertainty principles are truly sharper than the existing ones in the literature. The new proposals' applications in time-frequency and optical system analysis are also given.
\end{abstract}
\end{frontmatter}
\section{Introduction}\label{sec:1}
\indent Uncertainty principle plays an important role in harmonic analysis, quantum mechanics, and time-frequency analysis \cite{Zhang2017,Zhang20191,Zhang20192}. The classical $N$-dimensional Heisenberg's uncertainty principle is given by the inequality \cite{Folland1997,Hardin2018}
\begin{equation}\label{eq1}
\int_{\mathbb{R}^N}\left\lVert\mathbf{x}-\mathbf{a}\right\rVert^2|f(\mathbf{x})|^2\mathrm{d}\mathbf{x}\int_{\mathbb{R}^N}\left\lVert\mathbf{w}-\mathbf{b}\right\rVert^2\left|\widehat{f}(\mathbf{w})\right|^2\mathrm{d}\mathbf{w}\geq\frac{N^2}{16\pi^2}\lVert f\rVert_2^4
\end{equation}
for any $f(\mathbf{x})\in L^2(\mathbb{R}^N)$ equipped with a natural norm $\lVert\cdot\rVert_2=\left(\int_{\mathbb{R}^N}|\cdot(\mathbf{x})|^2\mathrm{d}\mathbf{x}\right)^{\frac{1}{2}}$ known as the $L^2$-norm, and any $\mathbf{a}=(a_1,a_2,\cdots,a_N),\mathbf{b}=(b_1,b_2,\cdots,b_N)\in\mathbb{R}^N$ equipped with the 2-norm $\lVert\cdot\rVert=\sqrt{(\cdot)(\cdot)^{\mathrm{T}}}$, where $\mathrm{T}$ denotes the transpose operator. The function $\widehat{f}(\mathbf{w})$ denotes the $N$-dimensional Fourier transform (FT) of $f(\mathbf{x})$ \cite{Osgood2014},
\begin{equation}\label{eq2}
\widehat{f}(\mathbf{w})=\int_{\mathbb{R}^N}f(\mathbf{x})e^{-2\pi i\mathbf{x}\mathbf{w}^{\mathrm{T}}}\mathrm{d}\mathbf{x},
\end{equation}
where $\mathbf{x}=(x_1,x_2,\cdots,x_N)$, $\mathbf{w}=(\omega_1,\omega_2,\cdots,\omega_N)$, and $\mathbf{x}\mathbf{w}^{\mathrm{T}}=\sum\limits_{k=1}^Nx_k\omega_k$. This version of uncertainty principle states that a multivariable square integrable function cannot be sharply localized in both the time domain and frequency domain. Given this result, it is theoretically important and practically useful to study its extension to the time-frequency domain.
\subsection{Overview and main result}\label{subsec1.1}
\indent In this paper we will focus on an interesting extension of the classical $N$-dimensional Heisenberg's uncertainty principle to the fractional Fourier transform (FRFT) \cite{Ozaktas2001}, which generalizes the FT by embedding another degree of freedom associated with rotational angle $\alpha$. Our main goal is to show that the classical result can be extended to time-frequency domain characterized by two $N$-dimensional FRFTs.\\
\indent Let us begin by recalling some background and notation on the $N$-dimensional FRFT \cite{Upadhyay2017}.
\begin{defn}\label{Def1}
The $N$-dimensional FRFT of a function $f(\mathbf{x})\in L^2(\mathbb{R}^N)$ with the rotational angle $\alpha$ is defined as
\begin{eqnarray}\label{eq3}
\mathcal{F}^{\alpha}[f](\mathbf{u})=\widehat{f}_{\alpha}(\mathbf{u})=\left\{
\begin{array}{ll}
\int_{\mathbb{R}^N}f(\mathbf{x})K_{\alpha}(\mathbf{x},\mathbf{u})\mathrm{d}\mathbf{x},&\alpha\neq n\pi\\
f(\mathbf{u}),&\alpha=2n\pi\\
f(-\mathbf{u}),&\alpha=(2n+1)\pi
\end{array}
\right.,n\in\mathbb{Z},
\end{eqnarray}
where $\mathbf{u}=(u_1,u_2,\cdots,u_N)$ and the kernel is
\begin{equation}\label{eq4}
K_{\alpha}(\mathbf{x},\mathbf{u})=(1-i\cot\alpha)^{\frac{N}{2}}e^{\pi i\left(\lVert\mathbf{x}\rVert^2+\lVert\mathbf{u}\rVert^2\right)\cot\alpha-2\pi i\mathbf{x}\mathbf{u}^{\mathrm{T}}\csc\alpha}.
\end{equation}
The corresponding inverse formula is given by $f(\mathbf{x})=\mathcal{F}^{-\alpha}\left[\widehat{f}_{\alpha}\right](\mathbf{x})$.
\end{defn}

\indent As it is seen, the $N$-dimensional FRFT of $\alpha=\left(2n+\frac{1}{2}\right)\pi,n\in\mathbb{Z}$ reduces to the $N$-dimensional FT. The fractional part of FRFT comes from the fact that another degree of freedom was added to the FT by introducing the parameter $\alpha$ which can be $\alpha\neq\left(n+\frac{1}{2}\right)\pi,n\pi,n\in\mathbb{Z}$. It is such a parameter that enables the FRFT to have flexibility to be used in scenarios that the FT is not applicable (non-stationary signal and image processing, time-frequency analysis, optical system analysis, etc.). There has been particular interest in FRFT's uncertainty principle since it can provide theoretical basis for many realistic applications, such as the effective bandwidth estimation and the quadratic phase system analysis. To be specific, the uncertainty principle in the FRFT domain was first investigated by Ozaktas et al. \cite{Ozaktas2001}. Shinde et al. proposed a stronger result on the uncertainty product in two FRFT domains for real functions \cite{Shinde2001}, and then Dang et al. extended this result to complex functions \cite{Dang2013}. In addition, Xu et al. discussed some extensions of Heisenberg's uncertainty principle on the FRFT, including the FRFT-based logarithmic, entropic and R\'{e}nyi entropic uncertainty principles \cite{Xu12009,Xu22009}. All of these results are dealing with single variable functions. However, in the literature, there are only a few scattered results on the high-dimensional case, see \cite{Zhang20192,Li2014} for related results. The one proposed in \cite{Li2014} for the $N$-dimensional FRFT is essentially the classical $N$-dimensional Heisenberg's uncertainty principle, and therefore, its lower bound is not the tightest. As for the latest one given by \cite{Zhang20192} for two $N$-dimensional FRFTs, its lower bound works only for real functions. The main contribution of this paper is to introduce a sharper lower bound on the uncertainty product for multivariable complex functions in two $N$-dimensional FRFT domains.\\
\indent We shall also need necessary background and notation on moments and spreads in time, frequency and FRFT domains, and the covariance and absolute covariance in order to give our main result.
\begin{defn}\label{Def2}
Let $\widehat{f}(\mathbf{w})$ be the $N$-dimensional FT of $f(\mathbf{x})=\lambda(\mathbf{x})e^{2\pi i\varphi(\mathbf{x})}\in L^2(\mathbb{R}^N)$, and $\widehat{f}_{\alpha}(\mathbf{u})$ be the $N$-dimensional FRFT of $f(\mathbf{x})$ with the rotational angle $\alpha$. Assume that for any $1\leq k\leq N$ the classical partial derivative $\frac{\partial\varphi}{\partial x_{k}}$ exists at any point $\mathbf{x}\in\mathbb{R}^N$, and $\mathbf{x}f(\mathbf{x}),\mathbf{w}\widehat{f}(\mathbf{w})\in L^2(\mathbb{R}^N)$. It is then well-defined that\\
\indent (i) The spread in the time domain:
\begin{equation}\label{eq5}
\vartriangle\mathbf{x}^2=\int_{\mathbb{R}^N}\left\lVert\mathbf{x}-\mathbf{x}^0\right\rVert^2|f(\mathbf{x})|^2\mathrm{d}\mathbf{x},
\end{equation}
where the moment vector in the time domain is
\begin{equation}\label{eq6}
\mathbf{x}^0=(x_1^0,x_2^0,\cdots,x_N^0),x_k^0=\int_{\mathbb{R}^N}x_k|f(\mathbf{x})|^2\mathrm{d}\mathbf{x}/\lVert f\rVert_2^2.
\end{equation}
\indent (ii) The spread in the frequency domain:
\begin{equation}\label{eq7}
\vartriangle\mathbf{w}^2=\int_{\mathbb{R}^N}\left\lVert\mathbf{w}-\mathbf{w}^0\right\rVert^2\left|\widehat{f}(\mathbf{w})\right|^2\mathrm{d}\mathbf{w},
\end{equation}
where the moment vector in the frequency domain is
\begin{equation}\label{eq8}
\mathbf{w}^0=(\omega_1^0,\omega_2^0,\cdots,\omega_N^0),\omega_k^0=\int_{\mathbb{R}^N}\omega_k\left|\widehat{f}(\mathbf{w})\right|^2\mathrm{d}\mathbf{w}/\lVert f\rVert_2^2.
\end{equation}
\indent (iii) The spread in the FRFT domain:
\begin{equation}\label{eq9}
\vartriangle\mathbf{u}_{\alpha}^2=\int_{\mathbb{R}^N}\left\lVert\mathbf{u}-\mathbf{u}^{\alpha,0}\right\rVert^2\left|\widehat{f}_{\alpha}(\mathbf{u})\right|^2\mathrm{d}\mathbf{u},
\end{equation}
where the moment vector in the FRFT domain is
\begin{equation}\label{eq10}
\mathbf{u}^{\alpha,0}=(u_1^{\alpha,0},u_2^{\alpha,0},\cdots,u_N^{\alpha,0}),u_k^{\alpha,0}=\int_{\mathbb{R}^N}u_k\left|\widehat{f}_{\alpha}(\mathbf{u})\right|^2\mathrm{d}\mathbf{u}/\lVert f\rVert_2^2.
\end{equation}
\indent (iv) The covariance and absolute covariance:
\begin{equation}\label{eq11}
\textmd{Cov}_{\mathbf{x},\mathbf{w}}=\int_{\mathbb{R}^N}\left(\mathbf{x}-\mathbf{x}^0\right)\left(\nabla_{\mathbf{x}}\varphi-\mathbf{w}^0\right)^{\mathrm{T}}\lambda^2(\mathbf{x})\mathrm{d}\mathbf{x}
\end{equation}
and
\begin{equation}\label{eq12}
\textmd{COV}_{\mathbf{x},\mathbf{w}}=\int_{\mathbb{R}^N}\left|\mathbf{x}-\mathbf{x}^0\right|\left|\nabla_{\mathbf{x}}\varphi-\mathbf{w}^0\right|^{\mathrm{T}}\lambda^2(\mathbf{x})\mathrm{d}\mathbf{x},
\end{equation}
where $\nabla_{\mathbf{x}}\varphi=\left(\frac{\partial\varphi}{\partial x_{1}},\frac{\partial\varphi}{\partial x_{2}},\cdots,\frac{\partial\varphi}{\partial x_{N}}\right)$ denotes the gradient vector of $\varphi$. Here an absolute operator is applied to vectors and we mean an element-wise absolute value. It should also be noted that there is an inequality $\textmd{COV}_{\mathbf{x},\mathbf{w}}\geq\textmd{Cov}_{\mathbf{x},\mathbf{w}}$.
\end{defn}

\indent Our main result is the following. This result presents an uncertainty principle associated with complex functions' uncertainty product in two $N$-dimensional FRFT domains.
\begin{thm}\label{Th1}
Let $\widehat{f}(\mathbf{w})$ be the $N$-dimensional FT of $f(\mathbf{x})=\lambda(\mathbf{x})e^{2\pi i\varphi(\mathbf{x})}\in L^2(\mathbb{R}^N)$, and $\widehat{f}_{\alpha}(\mathbf{u}),\widehat{f}_{\beta}(\mathbf{u})$ be the $N$-dimensional FRFTs of $f(\mathbf{x})$ with rotational angles $\alpha,\beta$ respectively. Assume that for any $1\leq k\leq N$ the classical partial derivatives $\frac{\partial\lambda}{\partial x_{k}},\frac{\partial\varphi}{\partial x_{k}},\frac{\partial f}{\partial x_{k}}$ exist at any point $\mathbf{x}\in\mathbb{R}^N$, and $\mathbf{x}f(\mathbf{x}),\mathbf{w}\widehat{f}(\mathbf{w})\in L^2(\mathbb{R}^N)$. Then,
\begin{eqnarray}\label{eq13}
\vartriangle\mathbf{u}_{\alpha}^2\vartriangle\mathbf{u}_{\beta}^2&\geq&\left(\frac{N^2}{16\pi^2}\lVert f\rVert_2^4+\textmd{COV}_{\mathbf{x},\mathbf{w}}^2-\textmd{Cov}_{\mathbf{x},\mathbf{w}}^2\right)\sin^2(\alpha-\beta)\nonumber\\
&&+\left[\cos\alpha\cos\beta\vartriangle\mathbf{x}^2+\sin\alpha\sin\beta\vartriangle\mathbf{w}^2+\sin(\alpha+\beta)\textmd{Cov}_{\mathbf{x},\mathbf{w}}\right]^2,
\end{eqnarray}
where $\vartriangle\mathbf{x}^2,\vartriangle\mathbf{w}^2,\vartriangle\mathbf{u}_{\alpha}^2,\vartriangle\mathbf{u}_{\beta}^2,\textmd{Cov}_{\mathbf{x},\mathbf{w}},\textmd{COV}_{\mathbf{x},\mathbf{w}}$ are defined as shown in Definition~\ref{Def2}. If $\nabla_{\mathbf{x}}\varphi$ is continuous and $\lambda$ is non-zero almost everywhere, then the equality holds if and only if $f(\mathbf{x})$ is a chirp function with the form
\begin{equation}\label{eq14}
f(\mathbf{x})=e^{-\frac{1}{2\zeta}\left\lVert\mathbf{x}-\mathbf{x}^0\right\rVert^2+d}e^{2\pi i\left[\frac{1}{2\varepsilon}\sum\limits_{m=1}^N\eta(x_m)\left(x_m-x_m^0\right)^2+\mathbf{w}^0\mathbf{x}^{\mathrm{T}}+d^{\eta(x_1),\eta(x_2),\cdots,\eta(x_N)}\right]}
\end{equation}
for some $\zeta,\varepsilon>0$ and $d,d^{\eta(x_1),\eta(x_2),\cdots,\eta(x_N)}\in\mathbb{R}$, where
\begin{eqnarray}\label{eq15}
\eta(x_m)=\left\{
\begin{array}{ll}
1,&m\in\mathbf{k}_{j_1}\\
-1,&m\in\mathbf{k}_{j_2}\\
\textrm{sgn}\left(x_m-x_m^0\right),&m\in\mathbf{k}_{j_3}\\
-\textrm{sgn}\left(x_m-x_m^0\right),&m\in\mathbf{k}_{j_4}
\end{array}
\right.,
\end{eqnarray}
and where
\begin{equation}\label{eq16}
\mathbf{k}_{j_1}=\left\{k_{11},k_{12},\cdots,k_{1j_1}\right\}=\left\{1\leq k\leq N|\frac{\partial\varphi}{\partial x_{k}}=\frac{1}{\varepsilon}\left(x_k-x_k^0\right)+\omega_k^0\right\},
\end{equation}
\begin{equation}\label{eq17}
\mathbf{k}_{j_2}=\left\{k_{21},k_{22},\cdots,k_{2j_2}\right\}=\left\{1\leq k\leq N|\frac{\partial\varphi}{\partial x_{k}}=-\frac{1}{\varepsilon}\left(x_k-x_k^0\right)+\omega_k^0\right\},
\end{equation}
\begin{eqnarray}\label{eq18}
\mathbf{k}_{j_3}=\left\{k_{31},k_{32},\cdots,k_{3j_3}\right\}=\left\{1\leq k\leq N|\frac{\partial\varphi}{\partial x_{k}}=\left\{
\begin{array}{ll}
\frac{1}{\varepsilon}\left(x_k-x_k^0\right)+\omega_k^0,&x_k\geq x_k^0\\
-\frac{1}{\varepsilon}\left(x_k-x_k^0\right)+\omega_k^0,&x_k<x_k^0
\end{array}
\right.\right\}
\end{eqnarray}
and
\begin{eqnarray}\label{eq19}
\mathbf{k}_{j_4}=\left\{k_{41},k_{42},\cdots,k_{4j_4}\right\}=\left\{1\leq k\leq N|\frac{\partial\varphi}{\partial x_{k}}=\left\{
\begin{array}{ll}
-\frac{1}{\varepsilon}\left(x_k-x_k^0\right)+\omega_k^0,&x_k\geq x_k^0\\
\frac{1}{\varepsilon}\left(x_k-x_k^0\right)+\omega_k^0,&x_k<x_k^0
\end{array}
\right.\right\}
\end{eqnarray}
satisfying $\bigcup\limits_{p=1}^{4}\mathbf{k}_{j_p}=\{1,2,\cdots,N\}$ and $\mathbf{k}_{j_p}\bigcap\mathbf{k}_{j_q}=\emptyset$ for $p\neq q$.
\end{thm}

\indent Inequality \eqref{eq13} of Theorem~\ref{Th1} gives a lower bound on the uncertainty product for multivariable complex functions in two $N$-dimensional FRFT domains. As it is seen, this result includes particular cases some well-known uncertainty inequalities, such as:\\
\indent (i) For $N=1$, it becomes the uncertainty inequality for one-dimensional FRFT introduced by Dang et al. \cite{Dang2013}.\\
\indent (ii) For $\beta=m\pi,m\in\mathbb{Z}$, it becomes
\begin{equation}\label{eq20}
\vartriangle\mathbf{x}^2\vartriangle\mathbf{u}_{\alpha}^2\geq\left(\frac{N^2}{16\pi^2}\lVert f\rVert_2^4+\textmd{COV}_{\mathbf{x},\mathbf{w}}^2-\textmd{Cov}_{\mathbf{x},\mathbf{w}}^2\right)\sin^2\alpha+\left[\cos\alpha\vartriangle\mathbf{x}^2+\sin\alpha\textmd{Cov}_{\mathbf{x},\mathbf{w}}\right]^2,
\end{equation}
which improves the uncertainty inequality for the $N$-dimensional FRFT proposed in \cite{Li2014}, i.e.,
\begin{equation}\label{eq21}
\vartriangle\mathbf{x}^2\vartriangle\mathbf{u}_{\alpha}^2\geq\frac{N^2}{16\pi^2}\lVert f\rVert_2^4\sin^2\alpha
\end{equation}
through providing a tighter lower bound.\\
\indent (iii) For $\textmd{Cov}_{\mathbf{x},\mathbf{w}}=0$ (e.g., real functions satisfying $\nabla_{\mathbf{x}}\varphi\equiv\mathbf{0}$), it becomes
\begin{equation}\label{eq22}
\vartriangle\mathbf{u}_{\alpha}^2\vartriangle\mathbf{u}_{\beta}^2\geq\left(\frac{N^2}{16\pi^2}\lVert f\rVert_2^4+\textmd{COV}_{\mathbf{x},\mathbf{w}}^2\right)\sin^2(\alpha-\beta)+\left[\cos\alpha\cos\beta\vartriangle\mathbf{x}^2+\sin\alpha\sin\beta\vartriangle\mathbf{w}^2\right]^2,
\end{equation}
which improves the uncertainty inequality for two $N$-dimensional FRFTs given by \cite{Zhang20192}, i.e.,
\begin{equation}\label{eq23}
\vartriangle\mathbf{u}_{\alpha}^2\vartriangle\mathbf{u}_{\beta}^2\geq\frac{N^2}{16\pi^2}\lVert f\rVert_2^4\sin^2(\alpha-\beta)+\left[\cos\alpha\cos\beta\vartriangle\mathbf{x}^2+\sin\alpha\sin\beta\vartriangle\mathbf{w}^2\right]^2
\end{equation}
through providing a tighter lower bound.\\
\indent Moreover, Theorem~\ref{Th1} of $\alpha=n\pi,\beta=\left(m+\frac{1}{2}\right)\pi,n,m\in\mathbb{Z}$ reduces to an uncertainty principle for $N$-dimensional FT given by the following corollary.
\begin{cor}\label{Cor1}
Let $\widehat{f}(\mathbf{w})$ be the $N$-dimensional FT of $f(\mathbf{x})=\lambda(\mathbf{x})e^{2\pi i\varphi(\mathbf{x})}\in L^2(\mathbb{R}^N)$. Assume that for any $1\leq k\leq N$ the classical partial derivatives $\frac{\partial\lambda}{\partial x_{k}},\frac{\partial\varphi}{\partial x_{k}},\frac{\partial f}{\partial x_{k}}$ exist at any point $\mathbf{x}\in\mathbb{R}^N$, and $\mathbf{x}f(\mathbf{x}),\mathbf{w}\widehat{f}(\mathbf{w})\in L^2(\mathbb{R}^N)$. Then,
\begin{eqnarray}\label{eq24}
&&\int_{\mathbb{R}^N}\left\lVert\mathbf{x}-\mathbf{x}^0\right\rVert^2|f(\mathbf{x})|^2\mathrm{d}\mathbf{x}\int_{\mathbb{R}^N}\left\lVert\mathbf{w}-\mathbf{w}^0\right\rVert^2\left|\widehat{f}(\mathbf{w})\right|^2\mathrm{d}\mathbf{w}\nonumber\\
&\geq&\frac{N^2}{16\pi^2}\lVert f\rVert_2^4+\left[\int_{\mathbb{R}^N}\left|\mathbf{x}-\mathbf{x}^0\right|\left|\nabla_{\mathbf{x}}\varphi-\mathbf{w}^0\right|^{\mathrm{T}}\lambda^2(\mathbf{x})\mathrm{d}\mathbf{x}\right]^2,
\end{eqnarray}
where $\mathbf{x}^0,\mathbf{w}^0$ are the moment vectors in time and frequency domains respectively, and $\nabla_{\mathbf{x}}\varphi$ is the gradient vector of $\varphi$. If $\nabla_{\mathbf{x}}\varphi$ is continuous and $\lambda$ is non-zero almost everywhere, then the equality holds if and only if $f(\mathbf{x})$ is a chirp function with the form \eqref{eq14}.
\end{cor}

\indent Inequality \eqref{eq24} of Corollary~\ref{Cor1} gives a lower bound on the product of a multivariable complex function's spread in time domain and that in frequency domain. In reality, the proof of Theorem~\ref{Th1} requires a main preparatory lemma proving that the moment vectors $\mathbf{x}^0,\mathbf{w}^0$ found in inequality \eqref{eq24} can be replaced by arbitrary $\mathbf{a},\mathbf{b}\in\mathbb{R}^N$. Thus this lemma can be stated as follows.
\begin{lem}\label{lem1}
Let $\widehat{f}(\mathbf{w})$ be the $N$-dimensional FT of $f(\mathbf{x})=\lambda(\mathbf{x})e^{2\pi i\varphi(\mathbf{x})}\in L^2(\mathbb{R}^N)$, and $\mathbf{a}=(a_1,a_2,\cdots,a_N),\mathbf{b}=(b_1,b_2,\cdots,b_N)\in\mathbb{R}^N$. Assume that for any $1\leq k\leq N$ the classical partial derivatives $\frac{\partial\lambda}{\partial x_{k}},\frac{\partial\varphi}{\partial x_{k}},\frac{\partial f}{\partial x_{k}}$ exist at any point $\mathbf{x}\in\mathbb{R}^N$, and $\mathbf{x}f(\mathbf{x}),\mathbf{w}\widehat{f}(\mathbf{w})\in L^2(\mathbb{R}^N)$. Then,
\begin{eqnarray}\label{eq25}
&&\int_{\mathbb{R}^N}\left\lVert\mathbf{x}-\mathbf{a}\right\rVert^2|f(\mathbf{x})|^2\mathrm{d}\mathbf{x}\int_{\mathbb{R}^N}\left\lVert\mathbf{w}-\mathbf{b}\right\rVert^2\left|\widehat{f}(\mathbf{w})\right|^2\mathrm{d}\mathbf{w}\nonumber\\
&\geq&\frac{N^2}{16\pi^2}\lVert f\rVert_2^4+\left[\int_{\mathbb{R}^N}\left|\mathbf{x}-\mathbf{a}\right|\left|\nabla_{\mathbf{x}}\varphi-\mathbf{b}\right|^{\mathrm{T}}\lambda^2(\mathbf{x})\mathrm{d}\mathbf{x}\right]^2,
\end{eqnarray}
where $\nabla_{\mathbf{x}}\varphi$ is the gradient vector of $\varphi$. If $\nabla_{\mathbf{x}}\varphi$ is continuous and $\lambda$ is non-zero almost everywhere, then the equality holds if and only if $f(\mathbf{x})$ is a chirp function with the form
\begin{equation}\label{eq26}
f(\mathbf{x})=e^{-\frac{1}{2\zeta}\left\lVert\mathbf{x}-\mathbf{a}\right\rVert^2+d}e^{2\pi i\left[\frac{1}{2\varepsilon}\sum\limits_{m=1}^N\eta(x_m)\left(x_m-a_m\right)^2+\mathbf{b}\mathbf{x}^{\mathrm{T}}+d^{\eta(x_1),\eta(x_2),\cdots,\eta(x_N)}\right]}
\end{equation}
for some $\zeta,\varepsilon>0$ and $d,d^{\eta(x_1),\eta(x_2),\cdots,\eta(x_N)}\in\mathbb{R}$, where
\begin{eqnarray}\label{eq27}
\eta(x_m)=\left\{
\begin{array}{ll}
1,&m\in\mathbf{k}_{j_1}\\
-1,&m\in\mathbf{k}_{j_2}\\
\textrm{sgn}\left(x_m-a_m\right),&m\in\mathbf{k}_{j_3}\\
-\textrm{sgn}\left(x_m-a_m\right),&m\in\mathbf{k}_{j_4}
\end{array}
\right.,
\end{eqnarray}
and where
\begin{equation}\label{eq28}
\mathbf{k}_{j_1}=\left\{k_{11},k_{12},\cdots,k_{1j_1}\right\}=\left\{1\leq k\leq N|\frac{\partial\varphi}{\partial x_{k}}=\frac{1}{\varepsilon}\left(x_k-a_k\right)+b_k\right\},
\end{equation}
\begin{equation}\label{eq29}
\mathbf{k}_{j_2}=\left\{k_{21},k_{22},\cdots,k_{2j_2}\right\}=\left\{1\leq k\leq N|\frac{\partial\varphi}{\partial x_{k}}=-\frac{1}{\varepsilon}\left(x_k-a_k\right)+b_k\right\},
\end{equation}
\begin{eqnarray}\label{eq30}
\mathbf{k}_{j_3}=\left\{k_{31},k_{32},\cdots,k_{3j_3}\right\}=\left\{1\leq k\leq N|\frac{\partial\varphi}{\partial x_{k}}=\left\{
\begin{array}{ll}
\frac{1}{\varepsilon}\left(x_k-a_k\right)+b_k,&x_k\geq a_k\\
-\frac{1}{\varepsilon}\left(x_k-a_k\right)+b_k,&x_k<a_k
\end{array}
\right.\right\}
\end{eqnarray}
and
\begin{eqnarray}\label{eq31}
\mathbf{k}_{j_4}=\left\{k_{41},k_{42},\cdots,k_{4j_4}\right\}=\left\{1\leq k\leq N|\frac{\partial\varphi}{\partial x_{k}}=\left\{
\begin{array}{ll}
-\frac{1}{\varepsilon}\left(x_k-a_k\right)+b_k,&x_k\geq a_k\\
\frac{1}{\varepsilon}\left(x_k-a_k\right)+b_k,&x_k<a_k
\end{array}
\right.\right\}
\end{eqnarray}
satisfying $\bigcup\limits_{p=1}^{4}\mathbf{k}_{j_p}=\{1,2,\cdots,N\}$ and $\mathbf{k}_{j_p}\bigcap\mathbf{k}_{j_q}=\emptyset$ for $p\neq q$.
\end{lem}

\indent Inequality \eqref{eq25} of Lemma~\ref{lem1} is a sharper $N$-dimensional Heisenberg's uncertainty inequality which improves the classical result \eqref{eq1} through providing a tighter lower bound.\\
\indent The remainder of this paper is structured as follows. Section~\ref{sec2} contains the proof of our main preparatory result, Lemma~\ref{lem1}. Section~\ref{sec3} contains the proof of our main result, Theorem~\ref{Th1}. To be specific, Section~\ref{subsec3.1} proves an important relation between spreads of a multivariable complex function in time, frequency and FRFT domains, and Section~\ref{subsec3.2} combines Lemma~\ref{lem1} with this relation to prove Theorem~\ref{Th1}. Section~\ref{sec4} presents example and experimental results. Potential applications are in Section~\ref{sec5}, and the conclusions follow in Section~\ref{sec6}.\\
\indent In the sequel, we denote by $\mathbb{R}$ the set of real numbers, by $\mathbb{R}^N$ the Cartesian product of $N$ real number collections, by $\mathbb{Z}$ the set of integers, by $\mathrm{T}$ the transpose operator, and by --- the complex conjugate operator. The 2-norm operator for vectors and $L^2$-norm operator for functions denote $\lVert\cdot\rVert=\sqrt{(\cdot)(\cdot)^{\mathrm{T}}}$ and $\lVert\cdot\rVert_2=\left(\int_{\mathbb{R}^N}|\cdot(\mathbf{x})|^2\mathrm{d}\mathbf{x}\right)^{\frac{1}{2}}$, respectively. The function has a complex form $f(\mathbf{x})=\lambda(\mathbf{x})e^{2\pi i\varphi(\mathbf{x})}$, unless we emphasize that it is real-valued. The notation $\vartriangle\mathbf{x}^2$, $\vartriangle\mathbf{w}^2$ and $\vartriangle\mathbf{u}_{\alpha}^2,\vartriangle\mathbf{u}_{\beta}^2$ denote spreads in time, frequency and FRFT domains respectively, the notation $\textmd{Cov}_{\mathbf{x},\mathbf{w}}$ and $\textmd{COV}_{\mathbf{x},\mathbf{w}}$ denote the covariance and absolute covariance respectively, the notation $\mathbf{x}^0$, $\mathbf{w}^0$ and $\mathbf{u}^{\alpha,0},\mathbf{u}^{\beta,0}$ denote moment vectors in time, frequency and FRFT domains respectively, and the notation $\nabla_{\mathbf{x}}\varphi$ denotes the gradient vector of $\varphi$. When an absolute operator is applied to vectors and we mean an element-wise absolute value.
\section{Proof of the main lemma}\label{sec2}
\indent This section gives the proof of our main preparatory result, Lemma~\ref{lem1} which is crucially needed in the proof of our main theorem.\\
\indent Lemma~\ref{lem1} presents a stronger Heisenberg's uncertainty principle for $N$-dimensional FT, as we discussed in Section~\ref{subsec1.1}. The proof of such an $N$-dimensional FT type of uncertainty principle involving the absolute covariance requires an additional preparatory result, Lemma~\ref{lem2}. We first state and prove this preparatory result, and then use it to prove Lemma~\ref{lem1} at the end of this section.\\
\indent Let us begin proofs by collecting Parseval's relations in $N$-dimensional FT and FRFT domains \cite{Osgood2014,Ozaktas2001}.\\
\indent(\textbf{Parseval's Relation}.) Let $\widehat{f}(\mathbf{w}),\widehat{g}(\mathbf{w})$ be the $N$-dimensional FTs of $f(\mathbf{x}),g(\mathbf{x})\in L^2(\mathbb{R}^N)$ respectively, and $\widehat{f}_{\alpha}(\mathbf{u}),\widehat{g}_{\alpha}(\mathbf{u})$ be the $N$-dimensional FRFTs of $f(\mathbf{x}),g(\mathbf{x})$ with the rotational angle $\alpha$ respectively, then
\begin{equation}\label{eq32}
\int_{\mathbb{R}^N}|f(\mathbf{x})|^2\mathrm{d}\mathbf{x}=\int_{\mathbb{R}^N}\left|\widehat{f}(\mathbf{w})\right|^2\mathrm{d}\mathbf{w}=\int_{\mathbb{R}^N}\left|\widehat{f}_{\alpha}(\mathbf{u})\right|^2\mathrm{d}\mathbf{u}
\end{equation}
and
\begin{equation}\label{eq33}
\int_{\mathbb{R}^N}f(\mathbf{x})\overline{g(\mathbf{x})}\mathrm{d}\mathbf{x}=\int_{\mathbb{R}^N}\widehat{f}(\mathbf{w})\overline{\widehat{g}(\mathbf{w})}\mathrm{d}\mathbf{w}=\int_{\mathbb{R}^N}\widehat{f}_{\alpha}(\mathbf{u})\overline{\widehat{g}_{\alpha}(\mathbf{u})}\mathrm{d}\mathbf{u}.
\end{equation}
\indent Here is the additional preparatory lemma.
\begin{lem}\label{lem2}
Let $\widehat{f}(\mathbf{w})$ be the $N$-dimensional FT of $f(\mathbf{x})=\lambda(\mathbf{x})e^{2\pi i\varphi(\mathbf{x})}\in L^2(\mathbb{R}^N)$, $b\in\mathbb{R}$, and $1\leq k\leq N$. Assume that the classical partial derivatives $\frac{\partial\lambda}{\partial x_{k}},\frac{\partial\varphi}{\partial x_{k}},\frac{\partial f}{\partial x_{k}}$ exist at any point $\mathbf{x}\in\mathbb{R}^N$, and $\omega_k\widehat{f}(\mathbf{w})\in L^2(\mathbb{R}^N)$. Then,
\begin{equation}\label{eq34}
\int_{\mathbb{R}^N}(\omega_k-b)^2\left|\widehat{f}(\mathbf{w})\right|^2\mathrm{d}\mathbf{w}=\frac{1}{4\pi^2}\int_{\mathbb{R}^N}\left(\frac{\partial\lambda}{\partial x_{k}}\right)^2\mathrm{d}\mathbf{x}+\int_{\mathbb{R}^N}\left(\frac{\partial\varphi}{\partial x_{k}}-b\right)^2\lambda^2(\mathbf{x})\mathrm{d}\mathbf{x}.
\end{equation}
\end{lem}
\begin{pro}
Using \eqref{eq32} of Parseval's relation in $N$-dimensional FT domain yields
\begin{equation}\label{eq35}
\int_{\mathbb{R}^N}(\omega_k-b)^2\left|\widehat{f}(\mathbf{w})\right|^2\mathrm{d}\mathbf{w}=\int_{\mathbb{R}^N}\omega_k^2\left|\widehat{f}(\mathbf{w})\right|^2\mathrm{d}\mathbf{w}+b^2\int_{\mathbb{R}^N}|f(\mathbf{x})|^2\mathrm{d}\mathbf{x}-2b\int_{\mathbb{R}^N}\omega_k\left|\widehat{f}(\mathbf{w})\right|^2\mathrm{d}\mathbf{w}.
\end{equation}
Since functions $\frac{1}{2\pi i}\frac{\partial f}{\partial x_{k}}$ and $\omega_k\widehat{f}(\mathbf{w})$ compose an $N$-dimensional FT pair, using \eqref{eq32} and \eqref{eq33} of Parseval's relation in $N$-dimensional FT domain, the above equation becomes
\begin{eqnarray}\label{eq36}
&&\int_{\mathbb{R}^N}(\omega_k-b)^2\left|\widehat{f}(\mathbf{w})\right|^2\mathrm{d}\mathbf{w}\nonumber\\
&=&\int_{\mathbb{R}^N}\left|\frac{1}{2\pi i}\frac{\partial f}{\partial x_{k}}\right|^2\mathrm{d}\mathbf{x}+b^2\int_{\mathbb{R}^N}|f(\mathbf{x})|^2\mathrm{d}\mathbf{x}-2b\int_{\mathbb{R}^N}\frac{1}{2\pi i}\frac{\partial f}{\partial x_{k}}\overline{f(\mathbf{x})}\mathrm{d}\mathbf{x}\nonumber\\
&=&\frac{1}{4\pi^2}\int_{\mathbb{R}^N}\left[\left(\frac{\partial\lambda}{\partial x_{k}}\right)^2+4\pi^2\left(\frac{\partial\varphi}{\partial x_{k}}\right)^2\lambda^2(\mathbf{x})\right]\mathrm{d}\mathbf{x}+b^2\int_{\mathbb{R}^N}\lambda^2(\mathbf{x})\mathrm{d}\mathbf{x}\nonumber\\
&&+\frac{bi}{\pi}\int_{\mathbb{R}^N}\frac{\partial\lambda}{\partial x_{k}}\lambda(\mathbf{x})\mathrm{d}\mathbf{x}-2b\int_{\mathbb{R}^N}\frac{\partial\varphi}{\partial x_{k}}\lambda^2(\mathbf{x})\mathrm{d}\mathbf{x}\nonumber\\
&=&\frac{1}{4\pi^2}\int_{\mathbb{R}^N}\left(\frac{\partial\lambda}{\partial x_{k}}\right)^2\mathrm{d}\mathbf{x}+\int_{\mathbb{R}^N}\left(\frac{\partial\varphi}{\partial x_{k}}-b\right)^2\lambda^2(\mathbf{x})\mathrm{d}\mathbf{x},
\end{eqnarray}
which gives the required result \eqref{eq34}.
\end{pro}

\indent We are now ready to prove Lemma~\ref{lem1}.
\begin{pro}[Proof of Lemma~\ref{lem1}]
It follows from \eqref{eq34} of Lemma~\ref{lem2} that for any $a,b\in\mathbb{R}$
\begin{equation}\label{eq37}
\int_{\mathbb{R}^N}(x_k-a)^2|f(\mathbf{x})|^2\mathrm{d}\mathbf{x}\int_{\mathbb{R}^N}(\omega_k-b)^2\left|\widehat{f}(\mathbf{w})\right|^2\mathrm{d}\mathbf{w}=\frac{1}{4\pi^2}I_1+I_2,
\end{equation}
where
\begin{equation}\label{eq38}
I_1=\int_{\mathbb{R}^N}(x_k-a)^2|f(\mathbf{x})|^2\mathrm{d}\mathbf{x}\int_{\mathbb{R}^N}\left(\frac{\partial\lambda}{\partial x_{k}}\right)^2\mathrm{d}\mathbf{x}
\end{equation}
and
\begin{equation}\label{eq39}
I_2=\int_{\mathbb{R}^N}(x_k-a)^2|f(\mathbf{x})|^2\mathrm{d}\mathbf{x}\int_{\mathbb{R}^N}\left(\frac{\partial\varphi}{\partial x_{k}}-b\right)^2\lambda^2(\mathbf{x})\mathrm{d}\mathbf{x}.
\end{equation}
Using the Cauchy-Schwarz inequality \cite{Aldaz2015} yields
\begin{equation}\label{eq40}
I_1\geq\left[\int_{\mathbb{R}^N}(x_k-a)\lambda(\mathbf{x})\frac{\partial\lambda}{\partial x_{k}}\mathrm{d}\mathbf{x}\right]^2=\left[\frac{1}{2}\int_{\mathbb{R}^N}|f(\mathbf{x})|^2\mathrm{d}\mathbf{x}\right]^2=\frac{\lVert f\rVert_2^4}{4}
\end{equation}
and
\begin{equation}\label{eq41}
I_2\geq\left[\int_{\mathbb{R}^N}\left|(x_k-a)\left(\frac{\partial\varphi}{\partial x_{k}}-b\right)\right|\lambda^2(\mathbf{x})\mathrm{d}\mathbf{x}\right]^2.
\end{equation}
With \eqref{eq37}, \eqref{eq40} and \eqref{eq41}, there is
\begin{eqnarray}\label{eq42}
&&\int_{\mathbb{R}^N}(x_k-a)^2|f(\mathbf{x})|^2\mathrm{d}\mathbf{x}\int_{\mathbb{R}^N}(\omega_k-b)^2\left|\widehat{f}(\mathbf{w})\right|^2\mathrm{d}\mathbf{w}\nonumber\\
&\geq&\frac{\lVert f\rVert_2^4}{16\pi^2}+\left[\int_{\mathbb{R}^N}\left|(x_k-a)\left(\frac{\partial\varphi}{\partial x_{k}}-b\right)\right|\lambda^2(\mathbf{x})\mathrm{d}\mathbf{x}\right]^2.
\end{eqnarray}
It follows from the Cauchy-Schwarz inequality \cite{Aldaz2015,Dragomir2003} that for any $\mathbf{a}=(a_1,a_2,\cdots,a_N),\mathbf{b}=(b_1,b_2,\cdots,b_N)\in\mathbb{R}^N$
\begin{eqnarray}\label{eq43}
&&\int_{\mathbb{R}^N}\left\lVert\mathbf{x}-\mathbf{a}\right\rVert^2|f(\mathbf{x})|^2\mathrm{d}\mathbf{x}\int_{\mathbb{R}^N}\left\lVert\mathbf{w}-\mathbf{b}\right\rVert^2\left|\widehat{f}(\mathbf{w})\right|^2\mathrm{d}\mathbf{w}\nonumber\\
&=&\sum\limits_{k=1}^N\int_{\mathbb{R}^N}\left(x_k-a_k\right)^2|f(\mathbf{x})|^2\mathrm{d}\mathbf{x}\sum\limits_{k=1}^N\int_{\mathbb{R}^N}\left(\omega_k-b_k\right)^2\left|\widehat{f}(\mathbf{w})\right|^2\mathrm{d}\mathbf{w}\nonumber\\
&\geq&\left[\sum\limits_{k=1}^N\left(\int_{\mathbb{R}^N}\left(x_k-a_k\right)^2|f(\mathbf{x})|^2\mathrm{d}\mathbf{x}\int_{\mathbb{R}^N}\left(\omega_k-b_k\right)^2\left|\widehat{f}(\mathbf{w})\right|^2\mathrm{d}\mathbf{w}\right)^{\frac{1}{2}}\right]^2.
\end{eqnarray}
Using \eqref{eq42} yields
\begin{eqnarray}\label{eq44}
&&\int_{\mathbb{R}^N}\left\lVert\mathbf{x}-\mathbf{a}\right\rVert^2|f(\mathbf{x})|^2\mathrm{d}\mathbf{x}\int_{\mathbb{R}^N}\left\lVert\mathbf{w}-\mathbf{b}\right\rVert^2\left|\widehat{f}(\mathbf{w})\right|^2\mathrm{d}\mathbf{w}\nonumber\\
&\geq&\left[\sum\limits_{k=1}^N\left(\frac{\lVert f\rVert_2^4}{16\pi^2}+\left(\int_{\mathbb{R}^N}\left|\left(x_k-a_k\right)\left(\frac{\partial\varphi}{\partial x_{k}}-b_k\right)\right|\lambda^2(\mathbf{x})\mathrm{d}\mathbf{x}\right)^2\right)^{\frac{1}{2}}\right]^2\nonumber\\
&=&\frac{\lVert f\rVert_2^4}{16\pi^2}\left[\sum\limits_{k=1}^N\left(1+\left(\frac{4\pi}{\lVert f\rVert_2^2}\int_{\mathbb{R}^N}\left|\left(x_k-a_k\right)\left(\frac{\partial\varphi}{\partial x_{k}}-b_k\right)\right|\lambda^2(\mathbf{x})\mathrm{d}\mathbf{x}\right)^2\right)^{\frac{1}{2}}\right]^2\nonumber\\
&\geq&\frac{\lVert f\rVert_2^4}{16\pi^2}\left[N^2+\left(\sum\limits_{k=1}^N\frac{4\pi}{\lVert f\rVert_2^2}\int_{\mathbb{R}^N}\left|\left(x_k-a_k\right)\left(\frac{\partial\varphi}{\partial x_{k}}-b_k\right)\right|\lambda^2(\mathbf{x})\mathrm{d}\mathbf{x}\right)^2\right]\nonumber\\
&=&\frac{N^2}{16\pi^2}\lVert f\rVert_2^4+\left[\int_{\mathbb{R}^N}\left|\mathbf{x}-\mathbf{a}\right|\left|\nabla_{\mathbf{x}}\varphi-\mathbf{b}\right|^{\mathrm{T}}\lambda^2(\mathbf{x})\mathrm{d}\mathbf{x}\right]^2,
\end{eqnarray}
which gives the required result \eqref{eq25}.\\
\indent Next we deduce the conditions under which the equality holds in \eqref{eq25}.\\
\indent The inequality \eqref{eq40} brings in conditions obeyed by the amplitude function $\lambda(\mathbf{x})$. The equality in \eqref{eq40} is attained if and only if there exists a positive number $\zeta$ such that
\begin{equation}\label{eq45}
(x_k-a)\lambda(\mathbf{x})=\zeta\frac{\partial\lambda}{\partial x_{k}}
\end{equation}
or
\begin{equation}\label{eq46}
-(x_k-a)\lambda(\mathbf{x})=\zeta\frac{\partial\lambda}{\partial x_{k}}.
\end{equation}
The first case shall not happen because it could result in a function $\lambda(\mathbf{x})\notin L^2(\mathbb{R}^N)$ \cite{Dang20131}. Then, $-\left(x_k-a_k\right)\lambda(\mathbf{x})=\zeta_k\frac{\partial\lambda}{\partial x_{k}}$ holds for all $1\leq k\leq N$ as the first equality in \eqref{eq44} holds. Solving the system of partial differential equations gives
\begin{equation}\label{eq47}
\lambda(\mathbf{x})=e^{\sum\limits_{k=1}^N-\frac{1}{2\zeta_k}\left(x_k-a_k\right)^2+d}.
\end{equation}
\indent The inequality \eqref{eq41} brings in conditions obeyed by the phase function $\varphi(\mathbf{x})$. The equality in \eqref{eq41} is attained if and only if there exists a positive number $\varepsilon$ such that
\begin{equation}\label{eq48}
\left|(x_k-a)\lambda(\mathbf{x})\right|=\varepsilon\left|\left(\frac{\partial\varphi}{\partial x_{k}}-b\right)\lambda(\mathbf{x})\right|,
\end{equation}
or equivalently,
\begin{equation}\label{eq49}
\left|x_k-a\right|=\varepsilon\left|\frac{\partial\varphi}{\partial x_{k}}-b\right|
\end{equation}
because of the almost everywhere non-zero of $\lambda$ and the continuity assumption of $\frac{\partial\varphi}{\partial x_{k}}$. Then, $\left|x_k-a_k\right|=\varepsilon_k\left|\frac{\partial\varphi}{\partial x_{k}}-b_k\right|$ holds for all $1\leq k\leq N$ as the first equality in \eqref{eq44} holds. As it is seen, there can be altogether four cases \cite{Dang20131}:
\begin{equation}\label{eq50}
\frac{\partial\varphi}{\partial x_{k}}=\frac{1}{\varepsilon_k}\left(x_k-a_k\right)+b_k,
\end{equation}
\begin{equation}\label{eq51}
\frac{\partial\varphi}{\partial x_{k}}=-\frac{1}{\varepsilon_k}\left(x_k-a_k\right)+b_k,
\end{equation}
\begin{eqnarray}\label{eq52}
\frac{\partial\varphi}{\partial x_{k}}=\left\{
\begin{array}{ll}
\frac{1}{\varepsilon_k}\left(x_k-a_k\right)+b_k,&x_k\geq a_k\\
-\frac{1}{\varepsilon_k}\left(x_k-a_k\right)+b_k,&x_k<a_k
\end{array}
\right.
\end{eqnarray}
and
\begin{eqnarray}\label{eq53}
\frac{\partial\varphi}{\partial x_{k}}=\left\{
\begin{array}{ll}
-\frac{1}{\varepsilon_k}\left(x_k-a_k\right)+b_k,&x_k\geq a_k\\
\frac{1}{\varepsilon_k}\left(x_k-a_k\right)+b_k,&x_k<a_k
\end{array}
\right.,
\end{eqnarray}
from which the set of $\{1,2,\cdots,N\}$ can be partitioned into the following four components:
\begin{equation}\label{eq54}
\mathbf{k}_{j_1}=\left\{k_{11},k_{12},\cdots,k_{1j_1}\right\}=\left\{1\leq k\leq N|\frac{\partial\varphi}{\partial x_{k}}=\frac{1}{\varepsilon_k}\left(x_k-a_k\right)+b_k\right\},
\end{equation}
\begin{equation}\label{eq55}
\mathbf{k}_{j_2}=\left\{k_{21},k_{22},\cdots,k_{2j_2}\right\}=\left\{1\leq k\leq N|\frac{\partial\varphi}{\partial x_{k}}=-\frac{1}{\varepsilon_k}\left(x_k-a_k\right)+b_k\right\},
\end{equation}
\begin{eqnarray}\label{eq56}
\mathbf{k}_{j_3}=\left\{k_{31},k_{32},\cdots,k_{3j_3}\right\}=\left\{1\leq k\leq N|\frac{\partial\varphi}{\partial x_{k}}=\left\{
\begin{array}{ll}
\frac{1}{\varepsilon_k}\left(x_k-a_k\right)+b_k,&x_k\geq a_k\\
-\frac{1}{\varepsilon_k}\left(x_k-a_k\right)+b_k,&x_k<a_k
\end{array}
\right.\right\}
\end{eqnarray}
and
\begin{eqnarray}\label{eq57}
\mathbf{k}_{j_4}=\left\{k_{41},k_{42},\cdots,k_{4j_4}\right\}=\left\{1\leq k\leq N|\frac{\partial\varphi}{\partial x_{k}}=\left\{
\begin{array}{ll}
-\frac{1}{\varepsilon_k}\left(x_k-a_k\right)+b_k,&x_k\geq a_k\\
\frac{1}{\varepsilon_k}\left(x_k-a_k\right)+b_k,&x_k<a_k
\end{array}
\right.\right\}.
\end{eqnarray}
Solving the system of partial differential equations yields
\begin{equation}\label{eq58}
\varphi(\mathbf{x})=\sum\limits_{m=1}^N\frac{1}{2\varepsilon_m}\eta(x_m)\left(x_m-a_m\right)^2+\mathbf{b}\mathbf{x}^{\mathrm{T}}+d^{\eta(x_1),\eta(x_2),\cdots,\eta(x_N)},
\end{equation}
where
\begin{eqnarray}\label{eq59}
\eta(x_m)=\left\{
\begin{array}{ll}
1,&m\in\mathbf{k}_{j_1}\\
-1,&m\in\mathbf{k}_{j_2}\\
\textrm{sgn}\left(x_m-a_m\right),&m\in\mathbf{k}_{j_3}\\
-\textrm{sgn}\left(x_m-a_m\right),&m\in\mathbf{k}_{j_4}
\end{array}
\right..
\end{eqnarray}
\indent The inequality \eqref{eq43} brings in conditions obeyed by the parameters $\zeta_k$ and $\varepsilon_k$, $k=1,2,\cdots,N$. The equality in \eqref{eq43} is attained if and only if the ratio
\begin{equation}\label{eq60}
\frac{\int_{\mathbb{R}^N}\left(x_k-a_k\right)^2|f(\mathbf{x})|^2\mathrm{d}\mathbf{x}}{\int_{\mathbb{R}^N}\left(\omega_k-b_k\right)^2\left|\widehat{f}(\mathbf{w})\right|^2\mathrm{d}\mathbf{w}}
\end{equation}
is a constant independent of $k$. Using \eqref{eq34}, \eqref{eq47} and \eqref{eq58}, it follows that
\begin{equation}\label{eq61}
\frac{1}{4\pi^2\zeta_k^2}+\frac{1}{\varepsilon_k^2}=\frac{1}{4\pi^2\zeta_l^2}+\frac{1}{\varepsilon_l^2}\;\textrm{for}\;k,l=1,2,\cdots,N.
\end{equation}
\indent The second inequality in \eqref{eq44} brings in conditions obeyed by the parameters $\zeta_k$ and $\varepsilon_k$, $k=1,2,\cdots,N$. It follows from \eqref{eq47} and \eqref{eq58} that
\begin{equation}\label{eq62}
\lVert f\rVert_2^2=e^{2d}\prod_{k=1}^{N}(\pi\zeta_k)^{\frac{1}{2}}
\end{equation}
and
\begin{equation}\label{eq63}
\int_{\mathbb{R}^N}\left|\left(x_k-a_k\right)\left(\frac{\partial\varphi}{\partial x_{k}}-b_k\right)\right|\lambda^2(\mathbf{x})\mathrm{d}\mathbf{x}=e^{2d}\frac{\zeta_k}{2\varepsilon_k}\prod_{k=1}^{N}(\pi\zeta_k)^{\frac{1}{2}},
\end{equation}
and then
\begin{equation}\label{eq64}
\frac{4\pi}{\lVert f\rVert_2^2}\int_{\mathbb{R}^N}\left|\left(x_k-a_k\right)\left(\frac{\partial\varphi}{\partial x_{k}}-b_k\right)\right|\lambda^2(\mathbf{x})\mathrm{d}\mathbf{x}=2\pi\frac{\zeta_k}{\varepsilon_k}.
\end{equation}
Thus, the second equality in \eqref{eq44} is attained if and only if the ratio
\begin{equation}\label{eq65}
\frac{\left[1+\left(2\pi\frac{\zeta_k}{\varepsilon_k}\right)^2\right]^{\frac{1}{2}}-1}{\left[1+\left(2\pi\frac{\zeta_k}{\varepsilon_k}\right)^2\right]^{\frac{1}{2}}+1}
\end{equation}
is a constant independent of $k$. It follows that
\begin{equation}\label{eq66}
\frac{\zeta_k}{\varepsilon_k}=\frac{\zeta_l}{\varepsilon_l}\;\textrm{for}\;k,l=1,2,\cdots,N.
\end{equation}
\indent With \eqref{eq61} and \eqref{eq66}, it concludes that $\zeta_k$ and $\varepsilon_k$ are constants independent of $k$, and denoted respectively by
\begin{equation}\label{eq67}
\zeta_k=\zeta,k=1,2,\cdots,N
\end{equation}
and
\begin{equation}\label{eq68}
\varepsilon_k=\varepsilon,k=1,2,\cdots,N.
\end{equation}
Then, the amplitude function \eqref{eq47} and the phase function \eqref{eq58} turn into
\begin{equation}\label{eq69}
\lambda(\mathbf{x})=e^{-\frac{1}{2\zeta}\left\lVert\mathbf{x}-\mathbf{a}\right\rVert^2+d}
\end{equation}
and
\begin{equation}\label{eq70}
\varphi(\mathbf{x})=\frac{1}{2\varepsilon}\sum\limits_{m=1}^N\eta(x_m)\left(x_m-a_m\right)^2+\mathbf{b}\mathbf{x}^{\mathrm{T}}+d^{\eta(x_1),\eta(x_2),\cdots,\eta(x_N)}
\end{equation}
respectively, giving rise to the required result \eqref{eq26}.
\end{pro}
\section{Proof of the main theorem}\label{sec3}
\indent This section gives the proof of our main result, Theorem~\ref{Th1}. In Section~\ref{subsec3.1}, we prove a technical lemma on the relationship between spreads of multivariable complex functions in time, frequency and FRFT domains. Then, in Section~\ref{subsec3.2} we combine this lemma with Lemma~\ref{lem1} to prove Theorem~\ref{Th1}.
\subsection{Relation between a multivariable complex function's spreads in time, frequency and FRFT domains}\label{subsec3.1}
\indent The spreads of a multivariable function $f(\mathbf{x})\in L^2(\mathbb{R}^N)$ in time, frequency and FRFT domains stand for its duration, bandwidth and FRFT-bandwidth, which are defined respectively as shown in \eqref{eq5}, \eqref{eq7} and \eqref{eq9} of Definition~\ref{Def2}. For a specific complex function $f(\mathbf{x})=\lambda(\mathbf{x})e^{2\pi i\varphi(\mathbf{x})}$, the relation between these spreads is given below.
\begin{lem}\label{lem3}
Let $\widehat{f}(\mathbf{w})$ be the $N$-dimensional FT of $f(\mathbf{x})=\lambda(\mathbf{x})e^{2\pi i\varphi(\mathbf{x})}\in L^2(\mathbb{R}^N)$, and $\widehat{f}_{\alpha}(\mathbf{u})$ be the $N$-dimensional FRFT of $f(\mathbf{x})$ with the rotational angle $\alpha$. Assume that for any $1\leq k\leq N$ the classical partial derivatives $\frac{\partial\lambda}{\partial x_{k}},\frac{\partial\varphi}{\partial x_{k}},\frac{\partial f}{\partial x_{k}}$ exist at any point $\mathbf{x}\in\mathbb{R}^N$, and $\mathbf{x}f(\mathbf{x}),\mathbf{w}\widehat{f}(\mathbf{w})\in L^2(\mathbb{R}^N)$. Then,
\begin{equation}\label{eq71}
\vartriangle\mathbf{u}_{\alpha}^2=\cos^2\alpha\vartriangle\mathbf{x}^2+\sin^2\alpha\vartriangle\mathbf{w}^2+2\sin\alpha\cos\alpha\textmd{Cov}_{\mathbf{x},\mathbf{w}},
\end{equation}
where $\vartriangle\mathbf{x}^2,\vartriangle\mathbf{w}^2,\vartriangle\mathbf{u}_{\alpha}^2,\textmd{Cov}_{\mathbf{x},\mathbf{w}}$ are defined as shown in Definition~\ref{Def2}.
\end{lem}
\begin{pro}
It follows from the definition of $N$-dimensional FRFT in the case of $\alpha=n\pi,n\in\mathbb{Z}$ that $\widehat{f}_{\alpha}(\mathbf{u})=f(\mathbf{u})$ or $\widehat{f}_{\alpha}(\mathbf{u})=f(-\mathbf{u})$, and then
\begin{equation}\label{eq72}
\vartriangle\mathbf{u}_{\alpha}^2=\int_{\mathbb{R}^N}\left\lVert\mathbf{u}-\mathbf{u}^{\alpha,0}\right\rVert^2\left|\widehat{f}_{\alpha}(\mathbf{u})\right|^2\mathrm{d}\mathbf{u}=\int_{\mathbb{R}^N}\left\lVert\mathbf{x}-\mathbf{x}^0\right\rVert^2|f(\mathbf{x})|^2\mathrm{d}\mathbf{x}=\vartriangle\mathbf{x}^2,
\end{equation}
which gives the required result \eqref{eq71} of $\alpha=n\pi,n\in\mathbb{Z}$. As for the case of $\alpha\neq n\pi,n\in\mathbb{Z}$, using \eqref{eq32} of Parseval's relation in $N$-dimensional FRFT domain gives for any $1\leq k\leq N$
\begin{equation}\label{eq73}
\int_{\mathbb{R}^N}\left(u_k-u_k^{\alpha,0}\right)^2\left|\widehat{f}_{\alpha}(\mathbf{u})\right|^2\mathrm{d}\mathbf{u}=\int_{\mathbb{R}^N}u_k^2\left|\widehat{f}_{\alpha}(\mathbf{u})\right|^2\mathrm{d}\mathbf{u}-\left(u_k^{\alpha,0}\right)^2\lVert f\rVert_2^2.
\end{equation}
In view of the $N$-dimensional FRFT's inverse formula, there is
\begin{equation}\label{eq74}
\frac{\sin\alpha}{2\pi i}\frac{\partial f}{\partial x_{k}}+\cos\alpha x_{k}f(\mathbf{x})=\int_{\mathbb{R}^N}u_k\widehat{f}_{\alpha}(\mathbf{u})K_{-\alpha}(\mathbf{u},\mathbf{x})\mathrm{d}\mathbf{u},
\end{equation}
which indicates that functions $\frac{\sin\alpha}{2\pi i}\frac{\partial f}{\partial x_{k}}+\cos\alpha x_{k}f(\mathbf{x})$ and $u_k\widehat{f}_{\alpha}(\mathbf{u})$ compose an $N$-dimensional FRFT pair. From \eqref{eq32} and \eqref{eq33} of Parseval's relation in $N$-dimensional FRFT domain, \eqref{eq73} becomes
\begin{eqnarray}\label{eq75}
\int_{\mathbb{R}^N}\left(u_k-u_k^{\alpha,0}\right)^2\left|\widehat{f}_{\alpha}(\mathbf{u})\right|^2\mathrm{d}\mathbf{u}&=&\int_{\mathbb{R}^N}\left|\frac{\sin\alpha}{2\pi i}\frac{\partial f}{\partial x_{k}}+\cos\alpha x_{k}f(\mathbf{x})\right|^2\mathrm{d}\mathbf{x}\nonumber\\
&&-\left[\int_{\mathbb{R}^N}\left(\frac{\sin\alpha}{2\pi i}\frac{\partial f}{\partial x_{k}}+\cos\alpha x_{k}f(\mathbf{x})\right)\overline{f(\mathbf{x})}\mathrm{d}\mathbf{x}\right]^2/\lVert f\rVert_2^2.
\end{eqnarray}
Because of \eqref{eq36}, the relations
\begin{equation}\label{eq76}
\int_{\mathbb{R}^N}\omega_k^2\left|\widehat{f}(\mathbf{w})\right|^2\mathrm{d}\mathbf{w}=\frac{1}{4\pi^2}\int_{\mathbb{R}^N}\left[\left(\frac{\partial\lambda}{\partial x_{k}}\right)^2+4\pi^2\left(\frac{\partial\varphi}{\partial x_{k}}\right)^2\lambda^2(\mathbf{x})\right]\mathrm{d}\mathbf{x}
\end{equation}
and
\begin{equation}\label{eq77}
\int_{\mathbb{R}^N}\omega_k\left|\widehat{f}(\mathbf{w})\right|^2\mathrm{d}\mathbf{w}=\int_{\mathbb{R}^N}\frac{\partial\varphi}{\partial x_{k}}\lambda^2(\mathbf{x})\mathrm{d}\mathbf{x}
\end{equation}
hold, resulting in
\begin{eqnarray}\label{eq78}
&&\int_{\mathbb{R}^N}\left|\frac{\sin\alpha}{2\pi i}\frac{\partial f}{\partial x_{k}}+\cos\alpha x_{k}f(\mathbf{x})\right|^2\mathrm{d}\mathbf{x}\nonumber\\
&=&\cos^2\alpha\int_{\mathbb{R}^N}x_k^2|f(\mathbf{x})|^2\mathrm{d}\mathbf{x}+\frac{\sin^2\alpha}{4\pi^2}\int_{\mathbb{R}^N}\left[\left(\frac{\partial\lambda}{\partial x_{k}}\right)^2+4\pi^2\left(\frac{\partial\varphi}{\partial x_{k}}\right)^2\lambda^2(\mathbf{x})\right]\mathrm{d}\mathbf{x}\nonumber\\
&&+2\sin\alpha\cos\alpha\int_{\mathbb{R}^N}x_k\frac{\partial\varphi}{\partial x_{k}}\lambda^2(\mathbf{x})\mathrm{d}\mathbf{x}\nonumber\\
&=&\cos^2\alpha\int_{\mathbb{R}^N}x_k^2|f(\mathbf{x})|^2\mathrm{d}\mathbf{x}+\sin^2\alpha\int_{\mathbb{R}^N}\omega_k^2\left|\widehat{f}(\mathbf{w})\right|^2\mathrm{d}\mathbf{w}\nonumber\\
&&+2\sin\alpha\cos\alpha\int_{\mathbb{R}^N}x_k\frac{\partial\varphi}{\partial x_{k}}\lambda^2(\mathbf{x})\mathrm{d}\mathbf{x}
\end{eqnarray}
and
\begin{eqnarray}\label{eq79}
\int_{\mathbb{R}^N}\left(\frac{\sin\alpha}{2\pi i}\frac{\partial f}{\partial x_{k}}+\cos\alpha x_{k}f(\mathbf{x})\right)\overline{f(\mathbf{x})}\mathrm{d}\mathbf{x}&=&\cos\alpha\int_{\mathbb{R}^N}x_k|f(\mathbf{x})|^2\mathrm{d}\mathbf{x}+\sin\alpha\int_{\mathbb{R}^N}\frac{\partial\varphi}{\partial x_{k}}\lambda^2(\mathbf{x})\mathrm{d}\mathbf{x}\nonumber\\
&=&\cos\alpha\int_{\mathbb{R}^N}x_k|f(\mathbf{x})|^2\mathrm{d}\mathbf{x}+\sin\alpha\int_{\mathbb{R}^N}\omega_k\left|\widehat{f}(\mathbf{w})\right|^2\mathrm{d}\mathbf{w}\nonumber\\
&=&\left(\cos\alpha x_k^0+\sin\alpha\omega_k^0\right)\lVert f\rVert_2^2.
\end{eqnarray}
Substituting into \eqref{eq75}, and using \eqref{eq32} of Parseval's relation in $N$-dimensional FT domain and \eqref{eq77} yields
\begin{eqnarray}\label{eq80}
&&\int_{\mathbb{R}^N}\left(u_k-u_k^{\alpha,0}\right)^2\left|\widehat{f}_{\alpha}(\mathbf{u})\right|^2\mathrm{d}\mathbf{u}\nonumber\\
&=&\cos^2\alpha\int_{\mathbb{R}^N}\left(x_k-x_k^0\right)^2|f(\mathbf{x})|^2\mathrm{d}\mathbf{x}+\sin^2\alpha\int_{\mathbb{R}^N}\left(\omega_k-\omega_k^0\right)^2\left|\widehat{f}(\mathbf{w})\right|^2\mathrm{d}\mathbf{w}\nonumber\\
&&+2\sin\alpha\cos\alpha\int_{\mathbb{R}^N}\left(x_k-x_k^0\right)\left(\frac{\partial\varphi}{\partial x_{k}}-\omega_k^0\right)\lambda^2(\mathbf{x})\mathrm{d}\mathbf{x},
\end{eqnarray}
and then
\begin{eqnarray}\label{eq81}
\vartriangle\mathbf{u}_{\alpha}^2&=&\int_{\mathbb{R}^N}\left\lVert\mathbf{u}-\mathbf{u}^{\alpha,0}\right\rVert^2\left|\widehat{f}_{\alpha}(\mathbf{u})\right|^2\mathrm{d}\mathbf{u}\nonumber\\
&=&\cos^2\alpha\int_{\mathbb{R}^N}\left\lVert\mathbf{x}-\mathbf{x}^0\right\rVert^2|f(\mathbf{x})|^2\mathrm{d}\mathbf{x}+\sin^2\alpha\int_{\mathbb{R}^N}\left\lVert\mathbf{w}-\mathbf{w}^0\right\rVert^2\left|\widehat{f}(\mathbf{w})\right|^2\mathrm{d}\mathbf{w}\nonumber\\
&&+2\sin\alpha\cos\alpha\int_{\mathbb{R}^N}\left(\mathbf{x}-\mathbf{x}^0\right)\left(\nabla_{\mathbf{x}}\varphi-\mathbf{w}^0\right)^{\mathrm{T}}\lambda^2(\mathbf{x})\mathrm{d}\mathbf{x}\nonumber\\
&=&\cos^2\alpha\vartriangle\mathbf{x}^2+\sin^2\alpha\vartriangle\mathbf{w}^2+2\sin\alpha\cos\alpha\textmd{Cov}_{\mathbf{x},\mathbf{w}}.
\end{eqnarray}
Combining \eqref{eq72} and \eqref{eq81} gives the required result \eqref{eq71}.
\end{pro}
\subsection{Combining Lemma~\ref{lem1} with the preparatory result: proof of the main theorem}\label{subsec3.2}
\indent In this section, we combine Lemma~\ref{lem3} with Lemma~\ref{lem1} to prove Theorem~\ref{Th1}.
\begin{pro}[Proof of Theorem~\ref{Th1}]
Using \eqref{eq71} of Lemma~\ref{lem3} gives
\begin{eqnarray}\label{eq82}
\vartriangle\mathbf{u}_{\alpha}^2\vartriangle\mathbf{u}_{\beta}^2&=&\left(\cos^2\alpha\vartriangle\mathbf{x}^2+\sin^2\alpha\vartriangle\mathbf{w}^2+2\sin\alpha\cos\alpha\textmd{Cov}_{\mathbf{x},\mathbf{w}}\right)\nonumber\\
&&\times\left(\cos^2\beta\vartriangle\mathbf{x}^2+\sin^2\beta\vartriangle\mathbf{w}^2+2\sin\beta\cos\beta\textmd{Cov}_{\mathbf{x},\mathbf{w}}\right)\nonumber\\
&=&\left(\vartriangle\mathbf{x}^2\vartriangle\mathbf{w}^2-\textmd{Cov}_{\mathbf{x},\mathbf{w}}^2\right)\sin^2(\alpha-\beta)\nonumber\\
&&+\left[\cos\alpha\cos\beta\vartriangle\mathbf{x}^2+\sin\alpha\sin\beta\vartriangle\mathbf{w}^2+\sin(\alpha+\beta)\textmd{Cov}_{\mathbf{x},\mathbf{w}}\right]^2.
\end{eqnarray}
Setting $\mathbf{a}=\mathbf{x}^0,\mathbf{b}=\mathbf{w}^0$ in \eqref{eq25} of Lemma~\ref{lem1}, it follows that \eqref{eq24} holds, i.e.,
\begin{equation}\label{eq83}
\vartriangle\mathbf{x}^2\vartriangle\mathbf{w}^2\geq\frac{N^2}{16\pi^2}\lVert f\rVert_2^4+\textmd{COV}_{\mathbf{x},\mathbf{w}}^2.
\end{equation}
Combining \eqref{eq82} with \eqref{eq83} yields
\begin{eqnarray}\label{eq84}
\vartriangle\mathbf{u}_{\alpha}^2\vartriangle\mathbf{u}_{\beta}^2&\geq&\left(\frac{N^2}{16\pi^2}\lVert f\rVert_2^4+\textmd{COV}_{\mathbf{x},\mathbf{w}}^2-\textmd{Cov}_{\mathbf{x},\mathbf{w}}^2\right)\sin^2(\alpha-\beta)\nonumber\\
&&+\left[\cos\alpha\cos\beta\vartriangle\mathbf{x}^2+\sin\alpha\sin\beta\vartriangle\mathbf{w}^2+\sin(\alpha+\beta)\textmd{Cov}_{\mathbf{x},\mathbf{w}}\right]^2,
\end{eqnarray}
which gives the required result \eqref{eq13}. As for the condition that reaches the equality relation, it is none other than the one giving rise to the equality in \eqref{eq83} (i.e., \eqref{eq24}). Thus the chirp function \eqref{eq26} of $\mathbf{a}=\mathbf{x}^0,\mathbf{b}=\mathbf{w}^0$ gives the required result \eqref{eq14}.
\end{pro}
\section{Example and numerical simulation}\label{sec4}
\indent In this section, we perform a two-dimensional example and simulation to illustrate the correctness of the derived results.\\
\indent Taking $N=2$ for example, the two-dimensional complex function is chosen as
\begin{equation}\label{eq85}
f(x_1,x_2)=e^{\sum\limits_{k=1}^2-\frac{1}{2\zeta_k}\left(x_k-x_k^0\right)^2+d}e^{2\pi i\left[\frac{1}{2\varepsilon_1}\left(x_1-x_1^0\right)^2-\frac{1}{2\varepsilon_2}\left(x_2-x_2^0\right)^2+\sum\limits_{m=1}^2\omega_m^0x_m+d_1\right]}
\end{equation}
that is a function of the amplitude form \eqref{eq47} and the phase form \eqref{eq58}, where $\zeta_k,\varepsilon_k>0$, $k=1,2$, $d,d_1\in\mathbb{R}$, and $e^{2d}\prod\limits_{k=1}^{2}(\pi\zeta_k)^{\frac{1}{2}}=1$. Then, it calculates that
\begin{equation}\label{eq86}
\lVert f\rVert_2^2=\int_{\mathbb{R}}\int_{\mathbb{R}}e^{2d}e^{\sum\limits_{k=1}^2-\frac{1}{\zeta_k}\left(x_k-x_k^0\right)^2}\mathrm{d}x_1\mathrm{d}x_2=e^{2d}\prod_{k=1}^2\int_{\mathbb{R}}e^{-\frac{1}{\zeta_k}x_k^2}\mathrm{d}x_k=e^{2d}\prod\limits_{k=1}^{2}(\pi\zeta_k)^{\frac{1}{2}}=1,
\end{equation}
\begin{eqnarray}\label{eq87}
\vartriangle\mathbf{x}^2&=&e^{2d}\int_{\mathbb{R}}\int_{\mathbb{R}}\left(x_1-x_1^0\right)^2e^{\sum\limits_{k=1}^2-\frac{1}{\zeta_k}\left(x_k-x_k^0\right)^2}\mathrm{d}x_1\mathrm{d}x_2\nonumber\\
&&+e^{2d}\int_{\mathbb{R}}\int_{\mathbb{R}}\left(x_2-x_2^0\right)^2e^{\sum\limits_{k=1}^2-\frac{1}{\zeta_k}\left(x_k-x_k^0\right)^2}\mathrm{d}x_1\mathrm{d}x_2\nonumber\\
&=&e^{2d}\int_{\mathbb{R}}x_1^2e^{-\frac{1}{\zeta_1}x_1^2}\mathrm{d}x_1\int_{\mathbb{R}}e^{-\frac{1}{\zeta_2}x_2^2}\mathrm{d}x_2+e^{2d}\int_{\mathbb{R}}x_2^2e^{-\frac{1}{\zeta_2}x_2^2}\mathrm{d}x_2\int_{\mathbb{R}}e^{-\frac{1}{\zeta_1}x_1^2}\mathrm{d}x_1\nonumber\\
&=&\frac{\zeta_1+\zeta_2}{2},
\end{eqnarray}
\begin{eqnarray}\label{eq88}
\vartriangle\mathbf{w}^2&=&\sum\limits_{k=1}^2\int_{\mathbb{R}}\int_{\mathbb{R}}\left(\omega_k-\omega_k^0\right)^2\left|\widehat{f}(\mathbf{w})\right|^2\mathrm{d}\mathbf{w}\nonumber\\
&=&\sum\limits_{k=1}^2\left[\frac{1}{4\pi^2}\int_{\mathbb{R}}\int_{\mathbb{R}}\left|\frac{\partial f}{\partial x_{k}}\right|^2\mathrm{d}x_1\mathrm{d}x_2-\left(\omega_k^0\right)^2\right]\nonumber\\
&=&\frac{1}{4\pi^2}\int_{\mathbb{R}}\int_{\mathbb{R}}\left[\frac{1}{\zeta_1^2}\left(x_1-x_1^0\right)^2+4\pi^2\left(\frac{1}{\varepsilon_1}\left(x_1-x_1^0\right)+\omega_1^0\right)^2\right]\nonumber\\
&&\relphantom{=====} {}\times e^{\sum\limits_{k=1}^2-\frac{1}{\zeta_k}\left(x_k-x_k^0\right)^2+2d}\mathrm{d}x_1\mathrm{d}x_2-\left(\omega_1^0\right)^2\nonumber\\
&&+\frac{1}{4\pi^2}\int_{\mathbb{R}}\int_{\mathbb{R}}\left[\frac{1}{\zeta_2^2}\left(x_2-x_2^0\right)^2+4\pi^2\left(-\frac{1}{\varepsilon_2}\left(x_2-x_2^0\right)+\omega_2^0\right)^2\right]\nonumber\\
&&\relphantom{======} {}\times e^{\sum\limits_{k=1}^2-\frac{1}{\zeta_k}\left(x_k-x_k^0\right)^2+2d}\mathrm{d}x_1\mathrm{d}x_2-\left(\omega_2^0\right)^2\nonumber\\
&=&\frac{\zeta_1}{2}\left(\frac{1}{4\pi^2\zeta_1^2}+\frac{1}{\varepsilon_1^2}\right)+\frac{\zeta_2}{2}\left(\frac{1}{4\pi^2\zeta_2^2}+\frac{1}{\varepsilon_2^2}\right),
\end{eqnarray}
\begin{eqnarray}\label{eq89}
\textmd{Cov}_{\mathbf{x},\mathbf{w}}&=&\sum\limits_{k=1}^2\int_{\mathbb{R}}\int_{\mathbb{R}}\left(x_k-x_k^0\right)\left(\frac{\partial\varphi}{\partial x_{k}}-\omega_k^0\right)e^{\sum\limits_{m=1}^2-\frac{1}{\zeta_m}\left(x_m-x_m^0\right)^2+2d}\mathrm{d}x_1\mathrm{d}x_2\nonumber\\
&=&\frac{1}{\varepsilon_1}\int_{\mathbb{R}}\int_{\mathbb{R}}\left(x_1-x_1^0\right)^2e^{\sum\limits_{m=1}^2-\frac{1}{\zeta_m}\left(x_m-x_m^0\right)^2+2d}\mathrm{d}x_1\mathrm{d}x_2\nonumber\\
&&-\frac{1}{\varepsilon_2}\int_{\mathbb{R}}\int_{\mathbb{R}}\left(x_2-x_2^0\right)^2e^{\sum\limits_{m=1}^2-\frac{1}{\zeta_m}\left(x_m-x_m^0\right)^2+2d}\mathrm{d}x_1\mathrm{d}x_2\nonumber\\
&=&\frac{\zeta_1}{2\varepsilon_1}-\frac{\zeta_2}{2\varepsilon_2},
\end{eqnarray}
\begin{eqnarray}\label{eq90}
\textmd{COV}_{\mathbf{x},\mathbf{w}}&=&\sum\limits_{k=1}^2\int_{\mathbb{R}}\int_{\mathbb{R}}\left|x_k-x_k^0\right|\left|\frac{\partial\varphi}{\partial x_{k}}-\omega_k^0\right|e^{\sum\limits_{m=1}^2-\frac{1}{\zeta_m}\left(x_m-x_m^0\right)^2+2d}\mathrm{d}x_1\mathrm{d}x_2\nonumber\\
&=&\sum\limits_{k=1}^2\frac{1}{\varepsilon_k}\int_{\mathbb{R}}\int_{\mathbb{R}}\left(x_k-x_k^0\right)^2e^{\sum\limits_{m=1}^2-\frac{1}{\zeta_m}\left(x_m-x_m^0\right)^2+2d}\mathrm{d}x_1\mathrm{d}x_2\nonumber\\
&=&\frac{\zeta_1}{2\varepsilon_1}+\frac{\zeta_2}{2\varepsilon_2},
\end{eqnarray}
\begin{eqnarray}\label{eq91}
\vartriangle\mathbf{u}_{\alpha}^2&=&\cos^2\alpha\vartriangle\mathbf{x}^2+\sin^2\alpha\vartriangle\mathbf{w}^2+2\sin\alpha\cos\alpha\textmd{Cov}_{\mathbf{x},\mathbf{w}}\nonumber\\
&=&\frac{\zeta_1+\zeta_2}{2}\cos^2\alpha+\left[\frac{\zeta_1}{2}\left(\frac{1}{4\pi^2\zeta_1^2}+\frac{1}{\varepsilon_1^2}\right)+\frac{\zeta_2}{2}\left(\frac{1}{4\pi^2\zeta_2^2}+\frac{1}{\varepsilon_2^2}\right)\right]\sin^2\alpha\nonumber\\
&&+\left(\frac{\zeta_1}{\varepsilon_1}-\frac{\zeta_2}{\varepsilon_2}\right)\sin\alpha\cos\alpha,
\end{eqnarray}
\begin{eqnarray}\label{eq92}
\vartriangle\mathbf{u}_{\beta}^2&=&\cos^2\beta\vartriangle\mathbf{x}^2+\sin^2\beta\vartriangle\mathbf{w}^2+2\sin\beta\cos\beta\textmd{Cov}_{\mathbf{x},\mathbf{w}}\nonumber\\
&=&\frac{\zeta_1+\zeta_2}{2}\cos^2\beta+\left[\frac{\zeta_1}{2}\left(\frac{1}{4\pi^2\zeta_1^2}+\frac{1}{\varepsilon_1^2}\right)+\frac{\zeta_2}{2}\left(\frac{1}{4\pi^2\zeta_2^2}+\frac{1}{\varepsilon_2^2}\right)\right]\sin^2\beta\nonumber\\
&&+\left(\frac{\zeta_1}{\varepsilon_1}-\frac{\zeta_2}{\varepsilon_2}\right)\sin\beta\cos\beta.
\end{eqnarray}
\indent From \eqref{eq87} and \eqref{eq88}, there is
\begin{eqnarray}\label{eq93}
\vartriangle\mathbf{x}^2\vartriangle\mathbf{w}^2&=&\frac{\zeta_1+\zeta_2}{2}\left[\frac{\zeta_1}{2}\left(\frac{1}{4\pi^2\zeta_1^2}+\frac{1}{\varepsilon_1^2}\right)+\frac{\zeta_2}{2}\left(\frac{1}{4\pi^2\zeta_2^2}+\frac{1}{\varepsilon_2^2}\right)\right]\nonumber\\
&=&\frac{1}{16\pi^2}\left(\frac{1}{\zeta_1}+\frac{1}{\zeta_2}\right)(\zeta_1+\zeta_2)+\frac{\zeta_1^2}{4\varepsilon_1^2}+\frac{\zeta_2^2}{4\varepsilon_2^2}+\left(\frac{1}{4\varepsilon_1^2}+\frac{1}{4\varepsilon_2^2}\right)\zeta_1\zeta_2.
\end{eqnarray}
Using the fact that the inequalities
\begin{equation}\label{eq94}
\frac{\zeta_2}{\zeta_1}+\frac{\zeta_1}{\zeta_2}\geq2
\end{equation}
and
\begin{equation}\label{eq95}
\frac{1}{\varepsilon_1^2}+\frac{1}{\varepsilon_2^2}\geq\frac{2}{\varepsilon_1\varepsilon_2}
\end{equation}
hold, it follows that
\begin{equation}\label{eq96}
\vartriangle\mathbf{x}^2\vartriangle\mathbf{w}^2\geq\frac{1}{4\pi^2}+\left(\frac{\zeta_1}{2\varepsilon_1}+\frac{\zeta_2}{2\varepsilon_2}\right)^2.
\end{equation}
It therefore concludes from \eqref{eq90} that
\begin{equation}\label{eq97}
\vartriangle\mathbf{x}^2\vartriangle\mathbf{w}^2\geq\frac{1}{4\pi^2}+\textmd{COV}_{\mathbf{x},\mathbf{w}}^2>\frac{1}{4\pi^2}.
\end{equation}
\indent From \eqref{eq87} and \eqref{eq91}, there is
\begin{eqnarray}\label{eq98}
\vartriangle\mathbf{x}^2\vartriangle\mathbf{u}_{\alpha}^2&=&\frac{(\zeta_1+\zeta_2)^2}{4}\cos^2\alpha+\frac{\zeta_1+\zeta_2}{2}\left[\frac{\zeta_1}{2}\left(\frac{1}{4\pi^2\zeta_1^2}+\frac{1}{\varepsilon_1^2}\right)+\frac{\zeta_2}{2}\left(\frac{1}{4\pi^2\zeta_2^2}+\frac{1}{\varepsilon_2^2}\right)\right]\sin^2\alpha\nonumber\\
&&+\frac{\zeta_1+\zeta_2}{2}\left(\frac{\zeta_1}{\varepsilon_1}-\frac{\zeta_2}{\varepsilon_2}\right)\sin\alpha\cos\alpha.
\end{eqnarray}
Using \eqref{eq96} gives
\begin{eqnarray}\label{eq99}
\vartriangle\mathbf{x}^2\vartriangle\mathbf{u}_{\alpha}^2&\geq&\frac{(\zeta_1+\zeta_2)^2}{4}\cos^2\alpha+\left[\frac{1}{4\pi^2}+\left(\frac{\zeta_1}{2\varepsilon_1}+\frac{\zeta_2}{2\varepsilon_2}\right)^2\right]\sin^2\alpha+\frac{\zeta_1+\zeta_2}{2}\left(\frac{\zeta_1}{\varepsilon_1}-\frac{\zeta_2}{\varepsilon_2}\right)\sin\alpha\cos\alpha\nonumber\\
&=&\left[\frac{1}{4\pi^2}+\left(\frac{\zeta_1}{2\varepsilon_1}+\frac{\zeta_2}{2\varepsilon_2}\right)^2-\left(\frac{\zeta_1}{2\varepsilon_1}-\frac{\zeta_2}{2\varepsilon_2}\right)^2\right]\sin^2\alpha\nonumber\\
&&+\left[\frac{\zeta_1+\zeta_2}{2}\cos\alpha+\left(\frac{\zeta_1}{2\varepsilon_1}-\frac{\zeta_2}{2\varepsilon_2}\right)\sin\alpha\right]^2.
\end{eqnarray}
It therefore concludes from \eqref{eq87}, \eqref{eq89} and \eqref{eq90} that
\begin{eqnarray}\label{eq100}
\vartriangle\mathbf{x}^2\vartriangle\mathbf{u}_{\alpha}^2&\geq&\left(\frac{1}{4\pi^2}+\textmd{COV}_{\mathbf{x},\mathbf{w}}^2-\textmd{Cov}_{\mathbf{x},\mathbf{w}}^2\right)\sin^2\alpha+\left[\cos\alpha\vartriangle\mathbf{x}^2+\sin\alpha\textmd{Cov}_{\mathbf{x},\mathbf{w}}\right]^2\nonumber\\
&>&\frac{1}{4\pi^2}\sin^2\alpha.
\end{eqnarray}
\indent From \eqref{eq91} and \eqref{eq92}, there is
\begin{eqnarray}\label{eq101}
&&\vartriangle\mathbf{u}_{\alpha}^2\vartriangle\mathbf{u}_{\beta}^2\nonumber\\
&=&\bigg[\frac{\zeta_1+\zeta_2}{2}\cos^2\alpha+\left(\frac{\zeta_1}{2}\left(\frac{1}{4\pi^2\zeta_1^2}+\frac{1}{\varepsilon_1^2}\right)+\frac{\zeta_2}{2}\left(\frac{1}{4\pi^2\zeta_2^2}+\frac{1}{\varepsilon_2^2}\right)\right)\sin^2\alpha+\left(\frac{\zeta_1}{\varepsilon_1}-\frac{\zeta_2}{\varepsilon_2}\right)\sin\alpha\cos\alpha\bigg]\nonumber\\
&&\times\bigg[\frac{\zeta_1+\zeta_2}{2}\cos^2\beta+\left(\frac{\zeta_1}{2}\left(\frac{1}{4\pi^2\zeta_1^2}+\frac{1}{\varepsilon_1^2}\right)+\frac{\zeta_2}{2}\left(\frac{1}{4\pi^2\zeta_2^2}+\frac{1}{\varepsilon_2^2}\right)\right)\sin^2\beta+\left(\frac{\zeta_1}{\varepsilon_1}-\frac{\zeta_2}{\varepsilon_2}\right)\sin\beta\cos\beta\bigg]\nonumber\\
&=&\left[\frac{\zeta_1+\zeta_2}{2}\left(\frac{\zeta_1}{2}\left(\frac{1}{4\pi^2\zeta_1^2}+\frac{1}{\varepsilon_1^2}\right)+\frac{\zeta_2}{2}\left(\frac{1}{4\pi^2\zeta_2^2}+\frac{1}{\varepsilon_2^2}\right)\right)-\left(\frac{\zeta_1}{2\varepsilon_1}-\frac{\zeta_2}{2\varepsilon_2}\right)^2\right]\sin^2(\alpha-\beta)\nonumber\\
&&+\Bigg[\frac{\zeta_1+\zeta_2}{2}\cos\alpha\cos\beta+\left(\frac{\zeta_1}{2}\left(\frac{1}{4\pi^2\zeta_1^2}+\frac{1}{\varepsilon_1^2}\right)+\frac{\zeta_2}{2}\left(\frac{1}{4\pi^2\zeta_2^2}+\frac{1}{\varepsilon_2^2}\right)\right)\sin\alpha\sin\beta\nonumber\\
&&\relphantom{==} {}+\left(\frac{\zeta_1}{2\varepsilon_1}-\frac{\zeta_2}{2\varepsilon_2}\right)\sin(\alpha+\beta)\Bigg]^2.
\end{eqnarray}
Using \eqref{eq96} yields
\begin{eqnarray}\label{eq102}
\vartriangle\mathbf{u}_{\alpha}^2\vartriangle\mathbf{u}_{\beta}^2&\geq&\left[\frac{1}{4\pi^2}+\left(\frac{\zeta_1}{2\varepsilon_1}+\frac{\zeta_2}{2\varepsilon_2}\right)^2-\left(\frac{\zeta_1}{2\varepsilon_1}-\frac{\zeta_2}{2\varepsilon_2}\right)^2\right]\sin^2(\alpha-\beta)\nonumber\\
&&+\bigg[\frac{\zeta_1+\zeta_2}{2}\cos\alpha\cos\beta+\left(\frac{\zeta_1}{2}\left(\frac{1}{4\pi^2\zeta_1^2}+\frac{1}{\varepsilon_1^2}\right)+\frac{\zeta_2}{2}\left(\frac{1}{4\pi^2\zeta_2^2}+\frac{1}{\varepsilon_2^2}\right)\right)\sin\alpha\sin\beta\nonumber\\
&&\relphantom{==} {}+\left(\frac{\zeta_1}{2\varepsilon_1}-\frac{\zeta_2}{2\varepsilon_2}\right)\sin(\alpha+\beta)\bigg]^2.
\end{eqnarray}
It therefore concludes from \eqref{eq87}--\eqref{eq90} that
\begin{eqnarray}\label{eq103}
\vartriangle\mathbf{u}_{\alpha}^2\vartriangle\mathbf{u}_{\beta}^2&\geq&\left(\frac{1}{4\pi^2}+\textmd{COV}_{\mathbf{x},\mathbf{w}}^2-\textmd{Cov}_{\mathbf{x},\mathbf{w}}^2\right)\sin^2(\alpha-\beta)\nonumber\\
&&+\left[\cos\alpha\cos\beta\vartriangle\mathbf{x}^2+\sin\alpha\sin\beta\vartriangle\mathbf{w}^2+\sin(\alpha+\beta)\textmd{Cov}_{\mathbf{x},\mathbf{w}}\right]^2.
\end{eqnarray}
Particularly, for $\frac{\zeta_1}{\varepsilon_1}=\frac{\zeta_2}{\varepsilon_2}$, i.e., $\textmd{Cov}_{\mathbf{x},\mathbf{w}}=0$, \eqref{eq103} becomes
\begin{eqnarray}\label{eq104}
\vartriangle\mathbf{u}_{\alpha}^2\vartriangle\mathbf{u}_{\beta}^2&\geq&\left(\frac{1}{4\pi^2}+\textmd{COV}_{\mathbf{x},\mathbf{w}}^2\right)\sin^2(\alpha-\beta)+\left[\cos\alpha\cos\beta\vartriangle\mathbf{x}^2+\sin\alpha\sin\beta\vartriangle\mathbf{w}^2\right]^2\nonumber\\
&>&\frac{1}{4\pi^2}\sin^2(\alpha-\beta)+\left[\cos\alpha\cos\beta\vartriangle\mathbf{x}^2+\sin\alpha\sin\beta\vartriangle\mathbf{w}^2\right]^2.
\end{eqnarray}
\indent When $\zeta_1=\zeta_2=\zeta$ and $\varepsilon_1=\varepsilon_2=\varepsilon$, the equality relations in \eqref{eq94} and \eqref{eq95} hold, implying that the equality in \eqref{eq96} is attained. It therefore concludes that
\begin{equation}\label{eq105}
\vartriangle\mathbf{x}^2\vartriangle\mathbf{w}^2=\frac{1}{4\pi^2}+\textmd{COV}_{\mathbf{x},\mathbf{w}}^2=\frac{1}{4\pi^2}+\frac{\zeta^2}{\varepsilon^2}>\frac{1}{4\pi^2},
\end{equation}
\begin{eqnarray}\label{eq106}
\vartriangle\mathbf{x}^2\vartriangle\mathbf{u}_{\alpha}^2&=&\left(\frac{1}{4\pi^2}+\textmd{COV}_{\mathbf{x},\mathbf{w}}^2-\textmd{Cov}_{\mathbf{x},\mathbf{w}}^2\right)\sin^2\alpha+\left[\cos\alpha\vartriangle\mathbf{x}^2+\sin\alpha\textmd{Cov}_{\mathbf{x},\mathbf{w}}\right]^2\nonumber\\
&=&\left(\frac{1}{4\pi^2}+\frac{\zeta^2}{\varepsilon^2}\right)\sin^2\alpha+\zeta^2\cos^2\alpha\nonumber\\
&>&\frac{1}{4\pi^2}\sin^2\alpha,
\end{eqnarray}
\begin{eqnarray}\label{eq107}
\vartriangle\mathbf{u}_{\alpha}^2\vartriangle\mathbf{u}_{\beta}^2&=&\left(\frac{1}{4\pi^2}+\textmd{COV}_{\mathbf{x},\mathbf{w}}^2-\textmd{Cov}_{\mathbf{x},\mathbf{w}}^2\right)\sin^2(\alpha-\beta)\nonumber\\
&&+\left[\cos\alpha\cos\beta\vartriangle\mathbf{x}^2+\sin\alpha\sin\beta\vartriangle\mathbf{w}^2+\sin(\alpha+\beta)\textmd{Cov}_{\mathbf{x},\mathbf{w}}\right]^2\nonumber\\
&=&\left(\frac{1}{4\pi^2}+\frac{\zeta^2}{\varepsilon^2}\right)\sin^2(\alpha-\beta)+\left[\zeta\cos\alpha\cos\beta+\zeta\left(\frac{1}{4\pi^2\zeta^2}+\frac{1}{\varepsilon^2}\right)\sin\alpha\sin\beta\right]^2\nonumber\\
&>&\frac{1}{4\pi^2}\sin^2(\alpha-\beta)+\left[\cos\alpha\cos\beta\vartriangle\mathbf{x}^2+\sin\alpha\sin\beta\vartriangle\mathbf{w}^2\right]^2\nonumber\\
&=&\frac{1}{4\pi^2}\sin^2(\alpha-\beta)+\left[\zeta\cos\alpha\cos\beta+\zeta\left(\frac{1}{4\pi^2\zeta^2}+\frac{1}{\varepsilon^2}\right)\sin\alpha\sin\beta\right]^2.
\end{eqnarray}
\indent In the example let $\zeta_1=1$, $\zeta_2=\frac{1}{2}$, $\varepsilon_1=2$, $\varepsilon_2=1$, $\alpha=\frac{2\pi}{3}$ and $\beta=\frac{\pi}{6}$, it calculates that
\begin{equation}\label{eq108}
\vartriangle\mathbf{x}^2\vartriangle\mathbf{w}^2\approx0.309746582899407,
\end{equation}
\begin{equation}\label{eq109}
\frac{1}{4\pi^2}+\textmd{COV}_{\mathbf{x},\mathbf{w}}^2\approx0.275330295910584,
\end{equation}
\begin{equation}\label{eq110}
\frac{1}{4\pi^2}\approx0.025330295910584,
\end{equation}
\begin{equation}\label{eq111}
\vartriangle\mathbf{x}^2\vartriangle\mathbf{u}_{\alpha}^2\approx0.372934937174556,
\end{equation}
\begin{equation}\label{eq112}
\left(\frac{1}{4\pi^2}+\textmd{COV}_{\mathbf{x},\mathbf{w}}^2-\textmd{Cov}_{\mathbf{x},\mathbf{w}}^2\right)\sin^2\alpha+\left[\cos\alpha\vartriangle\mathbf{x}^2+\sin\alpha\textmd{Cov}_{\mathbf{x},\mathbf{w}}\right]^2\approx0.347122721932938,
\end{equation}
\begin{equation}\label{eq113}
\frac{1}{4\pi^2}\sin^2\alpha\approx0.018997721932938,
\end{equation}
\begin{equation}\label{eq114}
\vartriangle\mathbf{u}_{\alpha}^2\vartriangle\mathbf{u}_{\beta}^2\approx0.331041346184749,
\end{equation}
\begin{eqnarray}\label{eq115}
&&\left(\frac{1}{4\pi^2}+\textmd{COV}_{\mathbf{x},\mathbf{w}}^2-\textmd{Cov}_{\mathbf{x},\mathbf{w}}^2\right)\sin^2(\alpha-\beta)\nonumber\\
&+&\left[\cos\alpha\cos\beta\vartriangle\mathbf{x}^2+\sin\alpha\sin\beta\vartriangle\mathbf{w}^2+\sin(\alpha+\beta)\textmd{Cov}_{\mathbf{x},\mathbf{w}}\right]^2\nonumber\\
&&\approx0.296625059195926,
\end{eqnarray}
\begin{equation}\label{eq116}
\frac{1}{4\pi^2}\sin^2(\alpha-\beta)+\left[\cos\alpha\cos\beta\vartriangle\mathbf{x}^2+\sin\alpha\sin\beta\vartriangle\mathbf{w}^2\right]^2\approx0.046625059195926,
\end{equation}
then it concludes from the view point of numerical simulation that the results \eqref{eq97}, \eqref{eq100}, \eqref{eq103} and \eqref{eq104} hold.\\
\indent In the example let $\zeta_1=\zeta_2=1$, $\varepsilon_1=\varepsilon_2=2$, $\alpha=\frac{2\pi}{3}$ and $\beta=\frac{\pi}{6}$, it calculates that
\begin{equation}\label{eq117}
\vartriangle\mathbf{x}^2\vartriangle\mathbf{w}^2\approx0.275330295910584,
\end{equation}
\begin{equation}\label{eq118}
\frac{1}{4\pi^2}+\textmd{COV}_{\mathbf{x},\mathbf{w}}^2\approx0.275330295910584,
\end{equation}
\begin{equation}\label{eq119}
\frac{1}{4\pi^2}\approx0.025330295910584,
\end{equation}
\begin{equation}\label{eq120}
\vartriangle\mathbf{x}^2\vartriangle\mathbf{u}_{\alpha}^2\approx0.456497721932938,
\end{equation}
\begin{equation}\label{eq121}
\left(\frac{1}{4\pi^2}+\textmd{COV}_{\mathbf{x},\mathbf{w}}^2-\textmd{Cov}_{\mathbf{x},\mathbf{w}}^2\right)\sin^2\alpha+\left[\cos\alpha\vartriangle\mathbf{x}^2+\sin\alpha\textmd{Cov}_{\mathbf{x},\mathbf{w}}\right]^2\approx0.456497721932938,
\end{equation}
\begin{equation}\label{eq122}
\frac{1}{4\pi^2}\sin^2\alpha\approx0.018997721932938,
\end{equation}
\begin{equation}\label{eq123}
\vartriangle\mathbf{u}_{\alpha}^2\vartriangle\mathbf{u}_{\beta}^2\approx0.373795204665280,
\end{equation}
\begin{eqnarray}\label{eq124}
&&\left(\frac{1}{4\pi^2}+\textmd{COV}_{\mathbf{x},\mathbf{w}}^2-\textmd{Cov}_{\mathbf{x},\mathbf{w}}^2\right)\sin^2(\alpha-\beta)\nonumber\\
&+&\left[\cos\alpha\cos\beta\vartriangle\mathbf{x}^2+\sin\alpha\sin\beta\vartriangle\mathbf{w}^2+\sin(\alpha+\beta)\textmd{Cov}_{\mathbf{x},\mathbf{w}}\right]^2\nonumber\\
&&\approx0.373795204665280,
\end{eqnarray}
\begin{equation}\label{eq125}
\frac{1}{4\pi^2}\sin^2(\alpha-\beta)+\left[\cos\alpha\cos\beta\vartriangle\mathbf{x}^2+\sin\alpha\sin\beta\vartriangle\mathbf{w}^2\right]^2\approx0.123795204665280,
\end{equation}
then it concludes from the view point of numerical simulation that the results \eqref{eq105}, \eqref{eq106} and \eqref{eq107} hold.
\section{Potential applications}\label{sec5}
\indent In the classical $N$-dimensional Heisenberg's uncertainty principle case the largest universal lower bound $\frac{N^2}{16\pi^2}\lVert f\rVert_2^4$ for all functions can be reached only if $\textmd{COV}_{\mathbf{x},\mathbf{w}}=0$. The proposed new corollary provides full characterization of the functions that make the equality relation hold in the uncertainty inequality, giving rise to a tighter lower bound $\frac{N^2}{16\pi^2}\lVert f\rVert_2^4+\textmd{COV}_{\mathbf{x},\mathbf{w}}^2$, which includes particular case the classical one when $\textmd{COV}_{\mathbf{x},\mathbf{w}}=0$. The philosophy of the $N$-dimensional FRFT based uncertainty principles is similar to that for the uncertainty principles in the classical setting. In the classical uncertainty principle in two $N$-dimensional FRFT domains case the largest universal lower bound for all functions is $\frac{N^2}{16\pi^2}\lVert f\rVert_2^4\sin^2(\alpha-\beta)$. Our previous work shows that a sharper lower bound can be $\frac{N^2}{16\pi^2}\lVert f\rVert_2^4\sin^2(\alpha-\beta)+\left[\cos\alpha\cos\beta\vartriangle\mathbf{x}^2+\sin\alpha\sin\beta\vartriangle\mathbf{w}^2\right]^2$, but this holds only for real functions. In our current work, the proposed new theorem gives a further larger lower bound $\left(\frac{N^2}{16\pi^2}\lVert f\rVert_2^4+\textmd{COV}_{\mathbf{x},\mathbf{w}}^2\right)\sin^2(\alpha-\beta)+\left[\cos\alpha\cos\beta\vartriangle\mathbf{x}^2+\sin\alpha\sin\beta\vartriangle\mathbf{w}^2\right]^2$ for real functions, a special form of the derived universal lower bound $\left(\frac{N^2}{16\pi^2}\lVert f\rVert_2^4+\textmd{COV}_{\mathbf{x},\mathbf{w}}^2-\textmd{Cov}_{\mathbf{x},\mathbf{w}}^2\right)\sin^2(\alpha-\beta)+\left[\cos\alpha\cos\beta\vartriangle\mathbf{x}^2+\sin\alpha\sin\beta\vartriangle\mathbf{w}^2+\sin(\alpha+\beta)\textmd{Cov}_{\mathbf{x},\mathbf{w}}\right]^2$ for complex functions. In such a way the new results present stronger uncertainty inequalities that imply the weaker ones, disclosing more information on the uncertainty products to be estimated. Thus, the new uncertainty principles could be able to process whatever practical application problems the old ones might be useful in solving, resulting in better performance.\\
\indent An alternative mathematical formulation of the classical $N$-dimensional Heisenberg's uncertainty principle is
\begin{equation}\label{eq126}
\frac{N^2}{16\pi^2}\lVert f\rVert_2^4=\min\left\{\vartriangle\mathbf{x}^2\vartriangle\mathbf{w}^2:\mathbf{x}f(\mathbf{x}),\mathbf{w}\widehat{f}(\mathbf{w})\in L^2(\mathbb{R}^N)\right\},
\end{equation}
where the minimum value $\frac{N^2}{16\pi^2}\lVert f\rVert_2^4$ of the uncertainty product $\vartriangle\mathbf{x}^2\vartriangle\mathbf{w}^2$ can be reached. For most functions this limit usually cannot be achieved, and the corresponding uncertainty product is actually larger than $\frac{N^2}{16\pi^2}\lVert f\rVert_2^4$. Our result indicates that a better estimate is $\frac{N^2}{16\pi^2}\lVert f\rVert_2^4+\textmd{COV}_{\mathbf{x},\mathbf{w}}^2$. Similarly, an alternative mathematical formulation of the classical uncertainty principle for the $N$-dimensional FRFT is
\begin{equation}\label{eq127}
\frac{N^2}{16\pi^2}\lVert f\rVert_2^4\sin^2\alpha=\min\left\{\vartriangle\mathbf{x}^2\vartriangle\mathbf{u}_{\alpha}^2:\mathbf{x}f(\mathbf{x}),\mathbf{w}\widehat{f}(\mathbf{w})\in L^2(\mathbb{R}^N)\right\},
\end{equation}
where the minimum value $\frac{N^2}{16\pi^2}\lVert f\rVert_2^4\sin^2\alpha$ of the uncertainty product $\vartriangle\mathbf{x}^2\vartriangle\mathbf{u}_{\alpha}^2$ can be reached. Our previous result provides a better estimate which says that the uncertainty product cannot be smaller than $\frac{N^2}{16\pi^2}\lVert f\rVert_2^4\sin^2\alpha+\cos^2\alpha\left(\vartriangle\mathbf{x}^2\right)^2$ for real functions. Our current result shows that, because of $\textmd{COV}_{\mathbf{x},\mathbf{w}}^2\geq\textmd{Cov}_{\mathbf{x},\mathbf{w}}^2$, a further better estimate is $\left(\frac{N^2}{16\pi^2}\lVert f\rVert_2^4+\textmd{COV}_{\mathbf{x},\mathbf{w}}^2-\textmd{Cov}_{\mathbf{x},\mathbf{w}}^2\right)\sin^2\alpha+\left[\cos\alpha\vartriangle\mathbf{x}^2+\sin\alpha\textmd{Cov}_{\mathbf{x},\mathbf{w}}\right]^2$. Uncertainty principles are suitable for the effective estimation of bandwidths. For instance, if $\vartriangle\mathbf{x}^2$ is known, it follows that
\begin{equation}\label{eq128}
\vartriangle\mathbf{w}^2\geq\frac{\frac{N^2}{16\pi^2}\lVert f\rVert_2^4+\textmd{COV}_{\mathbf{x},\mathbf{w}}^2}{\vartriangle\mathbf{x}^2}\geq\frac{\frac{N^2}{16\pi^2}\lVert f\rVert_2^4}{\vartriangle\mathbf{x}^2}
\end{equation}
and
\begin{eqnarray}\label{eq129}
\vartriangle\mathbf{u}_{\alpha}^2&\geq&\frac{\left(\frac{N^2}{16\pi^2}\lVert f\rVert_2^4+\textmd{COV}_{\mathbf{x},\mathbf{w}}^2-\textmd{Cov}_{\mathbf{x},\mathbf{w}}^2\right)\sin^2\alpha+\left[\cos\alpha\vartriangle\mathbf{x}^2+\sin\alpha\textmd{Cov}_{\mathbf{x},\mathbf{w}}\right]^2}{\vartriangle\mathbf{x}^2}\nonumber\\
&\geq&\frac{\frac{N^2}{16\pi^2}\lVert f\rVert_2^4\sin^2\alpha+\cos^2\alpha\left(\vartriangle\mathbf{x}^2\right)^2}{\vartriangle\mathbf{x}^2}\nonumber\\
&\geq&\frac{\frac{N^2}{16\pi^2}\lVert f\rVert_2^4\sin^2\alpha}{\vartriangle\mathbf{x}^2}.
\end{eqnarray}
Note that even the second term of the above inequality chains usually cannot be reached, except it is a chirp function given by \eqref{eq14} of Theorem~\ref{Th1}.\\
\indent The FRFT provides a mathematical model for analyzing and describing optical systems composed of an arbitrary sequence of thin lenses and sections of free space. Uncertainty relations are often used to estimate spreads in transformation domains. Therefore, uncertainty principles related to the spread in the FRFT domain reveal that the immediate application can be found in the discussion of some well-known optical physics phenomenons, such as the Fresnel diffraction and the FRFT system between planar surfaces \cite{Ozaktas1995}. We first consider a planar reference plane related to the scale parameter $s$. Fresnel diffracting it in the order observed at a distance $d$ from the screen can be described by an FRFT with the rotational angle $\alpha$ satisfying $\tan\alpha=\frac{d}{s^2}$. Using \eqref{eq20}, there exist an estimate to the spread of the observed field at the distance
\begin{equation}\label{eq130}
\vartriangle\mathbf{u}_{\alpha}^2\geq\frac{1}{1+\left(\frac{s^2}{d}\right)^2}\frac{\frac{N^2}{16\pi^2}\lVert f\rVert_2^4+\textmd{COV}_{\mathbf{x},\mathbf{w}}^2}{\vartriangle\mathbf{x}^2}+\frac{1}{1+\left(\frac{d}{s^2}\right)^2}\vartriangle\mathbf{x}^2+\frac{2}{\frac{d}{s^2}+\frac{s^2}{d}}\textmd{Cov}_{\mathbf{x},\mathbf{w}},
\end{equation}
implying that for short $d$ the effective spread $2\sqrt{\vartriangle\mathbf{u}_{\alpha}^2}$ is slightly larger than that at the planar reference plane $2\sqrt{\vartriangle\mathbf{x}^2}$ and for large distances $d$ the spread of the field is almost independent of $d$ and reciprocally proportional to the field spread in the planar reference plane. We then focus on two planar surfaces associated with the scale parameter $s$. Using a lens to compensate the spherical phase factors at both surfaces yields an FRFT system with the rotational angle $\alpha$ satisfying $\sin\alpha=\frac{d}{s^2}$ and $\tan\left(\frac{\alpha}{2}\right)=\frac{z}{s^2}$, where $d$ and $z$ denote the separation of the lenses and their focal length respectively. Therefore the relation \eqref{eq20} becomes
\begin{equation}\label{eq131}
\vartriangle\mathbf{u}_{\alpha}^2\geq\frac{d^2}{s^4}\frac{\frac{N^2}{16\pi^2}\lVert f\rVert_2^4+\textmd{COV}_{\mathbf{x},\mathbf{w}}^2}{\vartriangle\mathbf{x}^2}+\left(\frac{d}{z}-1\right)^2\vartriangle\mathbf{x}^2+2\frac{d}{s^2}\left(\frac{d}{z}-1\right)\textmd{Cov}_{\mathbf{x},\mathbf{w}},
\end{equation}
implying that for small $d$ the effective spread $2\sqrt{\vartriangle\mathbf{u}_{\alpha}^2}$ is slightly larger than that at the planar surfaces $2\sqrt{\vartriangle\mathbf{x}^2}$ and for a pair of $d$ and $z$ with similar values the spread of the field is proportional to $d$ or $z$ and reciprocally proportional to the field spread in the planar surfaces.
\section{Conclusions}\label{sec6}
\indent Uncertainty principles in two $N$-dimensional FRFT domains are investigated. The lower bounds obtained are tighter than the existing forms for three categories, those are, $N$-dimensional FT, $N$-dimensional FRFT and two $N$-dimensional FRFTs, in the literature. It turns out that the lower bounds are attainable by a chirp function with Gaussian envelop and quadratic phase. The correctness of the derived results is validated by example and experiment, and the effectiveness is illustrated by applications in the effective estimation of bandwidths in time-frequency analysis and spreads in optical system analysis.
\section*{Acknowledgments}
\indent The research was supported by the Startup Foundation for Introducing Talent of NUIST (Grant 2019r024).
\section*{References}
\bibliography{mybib}{}

\begin{thebibliography}{10}

\bibitem{Zhang2017}
Z.~C. Zhang.
\newblock Tighter uncertainty principles for linear canonical transform in
  terms of matrix decomposition.
\newblock {\em Digit. Signal Process.}, 69(10):70--85, October 2017.

\bibitem{Zhang20191}
Z.~C. Zhang.
\newblock Uncertainty principle for linear canonical transform using matrix
  decomposition of absolute spread matrix.
\newblock {\em Digit. Signal Process.}, 89(6):145--154, June 2019.

\bibitem{Zhang20192}
Z.~C. Zhang.
\newblock Uncertainty principle for real functions in free metaplectic
  transformation domains.
\newblock {\em J. Fourier Anal. Appl.}, 0(0):in press, May 2019.

\bibitem{Folland1997}
G.~B. Folland and A.~Sitaram.
\newblock The uncertainty principle: A mathematical survey.
\newblock {\em J. Fourier Anal. Appl.}, 3(3):207--238, May 1997.

\bibitem{Hardin2018}
D.~P. Hardin, M.~C.~Northington V, and A.~M. Powell.
\newblock A sharp balian-low uncertainty principle for shift-invariant spaces.
\newblock {\em Appl. Comput. Harmon. Anal.}, 44(2):294--311, March 2018.

\bibitem{Osgood2014}
B.~Osgood.
\newblock {\em Lecture Notes for EE 261 The Fourier Transform and its
  Applications}.
\newblock CreateSpace Independent Publishing Platform, 2014.

\bibitem{Ozaktas2001}
H.~M. Ozaktas, Z.~Zalevsky, and M.~A. Kutay.
\newblock {\em The Fractional Fourier Transform with Applications in Optics and
  Signal Processing}.
\newblock Wiley, New York, 2001.

\bibitem{Upadhyay2017}
S.~K. Upadhyay and J.~K. Dubey.
\newblock Wavelet convolution product involving fractional fourier transform.
\newblock {\em Fract. Calc. Appl. Anal.}, 20(1):173--189, February 2017.

\bibitem{Shinde2001}
S.~Shinde and V.~M. Gadre.
\newblock An uncertainty principle for real signals in the fractional fourier
  transform domain.
\newblock {\em IEEE Trans. Signal Process.}, 49(11):2545--2548, November 2001.

\bibitem{Dang2013}
P.~Dang, G.~T. Deng, and T.~Qian.
\newblock A tighter uncertainty principle for linear canonical transform in
  terms of phase derivative.
\newblock {\em IEEE Trans. Signal Process.}, 61(21):5153--5164, November 2013.

\bibitem{Xu12009}
G.~L. Xu, X.~T. Wang, and X.~G. Xu.
\newblock The logarithmic, heisenberg's and short-time uncertainty principles
  associated with fractional fourier transform.
\newblock {\em Signal Process.}, 89(3):339--343, March 2009.

\bibitem{Xu22009}
G.~L. Xu, X.~T. Wang, and X.~G. Xu.
\newblock Generalized entropic uncertainty principle on fractional fourier
  transform.
\newblock {\em Signal Process.}, 89(12):2692--2697, December 2009.

\bibitem{Li2014}
Y.~G. Li, B.~Z. Li, and H.~F. Sun.
\newblock Uncertainty principles for wigner-ville distribution associated with
  the linear canonical transforms.
\newblock {\em Abstr. Appl. Anal.}, 2014(470459):1--9, May 2014.

\bibitem{Aldaz2015}
J.~M. Aldaz, S.~Barza, M.~Fujii, and M.~S. Moslehian.
\newblock Advances in operator cauchy-schwarz inequalities and their reverses.
\newblock {\em Ann. Funct. Anal.}, 6(3):275--295, April 2015.

\bibitem{Dragomir2003}
S.~S. Dragomir.
\newblock A survey on cauchy-bunyakovsky-schwarz type discrete inequalities.
\newblock {\em J. Inequal. Pure Appl. Math.}, 4(3):1--142, May 2003.

\bibitem{Dang20131}
P.~Dang, G.~T. Deng, and T.~Qian.
\newblock A sharper uncertainty principle.
\newblock {\em J. Funct. Anal.}, 265(10):2239--2266, November 2013.

\bibitem{Ozaktas1995}
H.~M. Ozaktas and D.~Mendlovic.
\newblock Fractional fourier optics.
\newblock {\em J. Opt. Soc. Am. A}, 12(4):743--751, April 1995.

\end{thebibliography}
\bibliographystyle{unsrt}
\end{document}